\documentstyle[12pt]{article}
\begin{document}

\def\ds{\displaystyle}
\def\beq{\begin{equation}}
\def\eeq{\end{equation}}
\def\bea{\begin{eqnarray}}
\def\eea{\end{eqnarray}}
\def\beeq{\begin{eqnarray}}
\def\eeeq{\end{eqnarray}}
\def\ve{\vert}
\def\vel{\left|}
\def\ver{\right|}
\def\nnb{\nonumber}
\def\ga{\left(}
\def\dr{\right)}
\def\aga{\left\{}
\def\adr{\right\}}
\def\lla{\left<}
\def\rra{\right>}
\def\rar{\rightarrow}
\def\nnb{\nonumber}
\def\la{\langle}
\def\ra{\rangle}
\def\ba{\begin{array}}
\def\ea{\end{array}}
\def\tr{\mbox{Tr}}
\def\ssp{{\Sigma^{*+}}}
\def\sso{{\Sigma^{*0}}}
\def\ssm{{\Sigma^{*-}}}
\def\xis0{{\Xi^{*0}}}
\def\xism{{\Xi^{*-}}}
\def\qs{\la \bar s s \ra}
\def\qu{\la \bar u u \ra}
\def\qd{\la \bar d d \ra}
\def\qq{\la \bar q q \ra}
\def\gGgG{\la g^2 G^2 \ra}
\def\q{\gamma_5 \not\!q}
\def\x{\gamma_5 \not\!x}
\def\g5{\gamma_5}
\def\sb{S_Q^{cf}}
\def\sd{S_d^{be}}
\def\su{S_u^{ad}}
\def\ss{S_s^{??}}
\def\sbp{{S}_Q^{'cf}}
\def\sdp{{S}_d^{'be}}
\def\sup{{S}_u^{'ad}}
\def\ssp{{S}_s^{'??}}
\def\sig{\sigma_{\mu \nu} \gamma_5 p^\mu q^\nu}
\def\fo{f_0(\frac{s_0}{M^2})}
\def\ffi{f_1(\frac{s_0}{M^2})}
\def\fii{f_2(\frac{s_0}{M^2})}
\def\O{{\cal O}}
\def\sl{{\Sigma^0 \Lambda}}
\def\es{\!\!\! &=& \!\!\!}
\def\ap{\!\!\! &\approx& \!\!\!}
\def\ar{&+& \!\!\!}
\def\ek{&-& \!\!\!}
\def\kek{\!\!\!&-& \!\!\!}
\def\cp{&\times& \!\!\!}
\def\se{\!\!\! &\simeq& \!\!\!}
\def\eqv{&\equiv& \!\!\!}
\def\kpm{&\pm& \!\!\!}
\def\kmp{&\mp& \!\!\!}


\def\simlt{\stackrel{<}{{}_\sim}}
\def\simgt{\stackrel{>}{{}_\sim}}


\title{
         {\Large
                 {\bf
Polarized lepton pair forward--backward asymmetries in 
$B \rar K^\ast \ell^+ \ell^-$ decay beyond the standard model 
                 }
         }
      }

\author{\vspace{1cm}\\
{\small T. M. Aliev \thanks
{e-mail: taliev@metu.edu.tr}\,\,,
V. Bashiry
\,\,,
M. Savc{\i} \thanks
{e-mail: savci@metu.edu.tr}} \\
{\small Physics Department, Middle East Technical University,
06531 Ankara, Turkey} }

\date{}

\begin{titlepage}
\maketitle
\thispagestyle{empty}

\begin{abstract}
We study the polarized lepton pair forward--backward asymmetries in $B \rar
K^\ast \ell^+ \ell^-$ decay using a general, model independent form
of the effective Hamiltonian. We present the general expression for nine
double--polarization forward--backward asymmetries. It is shown that, the
study of the forward--backward asymmetries of the doubly--polarized lepton 
pair is a very useful tool for establishing new physics beyond the standard
model.
\end{abstract}

~~~PACS numbers: 13.20.He, 12.60.--i, 13.88.+e
\end{titlepage}

\section{Introduction}
Rare $B$ meson decays, induced by flavor changing neutral current (FCNC)
$b \rar s(d) \ell^+ \ell^-$ transitions provide a promising ground for
testing the structure of weak interactions. These decays which are forbidden 
in the standard model (SM) at tree level, occur at loop level and are very
sensitive to the gauge structure of the SM. Moreover, these decays are also 
quite sensitive to the existence of new physics beyond the SM, since loops
with new particles can give considerable contribution to rare decays. The
new physics effects in rare decays can appear in two ways; one via
modification of the existing Wilson coefficients in the SM, or through the
introduction of some operators with new coefficients. Theoretical
investigation of the $B \rar X_s \ell^+ \ell^-$ decays are relatively
more clean compared to their exclusive counterparts, since they are not
spoiled by nonperturbative long distance effects, while the corresponding 
exclusive channels are easier to measure experimentally.
Some of the most important exclusive FCNC decays are $B \rar K^\ast \gamma$
and $B \rar (\pi,\rho,K,K^\ast) \ell^+ \ell^-$ decays. The latter provides
potentially a very rich set of experimental observables, such as, lepton
pair forward--backward (FB) asymmetry, lepton polarizations, etc. 
Various kinematical distributions of such processes as
$B \rar K (K^\ast) \ell^+ \ell^-$ \cite{R6301,R6302,R6303}, 
$B \rar \pi (\rho) \ell^+ \ell^-$ \cite{R6304}, 
$B_{s,d} \rar \ell^+ \ell^-$ \cite{R6305} and 
$B_{s,d} \rar \gamma \ell^+ \ell^-$ \cite{R6306} have already been 
studied. Experimentally measurable quantities such as forward--backward
asymmetry, single polarization asymmetry, etc., have been studied for the 
$B \rar K^\ast \ell^+ \ell^-$ decay in \cite{R6301,R6307,R6308,R6309}. Study
of these quantities can give useful information in fitting the parameters of
the SM and put constraints on new physics \cite{R6310,R6311,R6312}. It has
been pointed out in \cite{R6313} that the study of simultaneous
polarizations of both leptons in the final state provide, in principle,
measurement of many more observables which would be useful in further
improvement of the parameters of the SM probing new physics beyond the SM.
It should be noted here that both lepton polarizations in the $B \rar K^\ast
\tau^+ \tau^-$ and $B \rar K \ell^+ \ell^-$ decays are studied in
\cite{R6314} and  \cite{R6315}, respectively.     
As has already been noted, one efficient way of establishing new physics
effects is studying forward--backward asymmetry in semileptonic $B \rar
K^\ast \ell^+ \ell^-$ decay, since, ${\cal A}_{FB}$ vanishes at specific
values of the dilepton invariant mass, and more essential than that,
this zero position of ${\cal A}_{FB}$ is known to be practically 
free of hadronic uncertainties \cite{R6312}.

The aim of the present work is studying the polarized forward--backward
asymmetry in the exclusive $B \rar K^\ast \ell^+ \ell^-$ decay using
a general form of the effective Hamiltonian, including all possible forms
of interactions. Here we would like to remind the reader that the influence
of new Wilson coefficients on various kinematical variables, such as branching
ratios, lepton pair forward--backward asymmetries and single lepton
polarization asymmetries for the inclusive $B \rar X_{s(d)} \ell^+ \ell^-$
decays (see first references in \cite{R6311,R6313,R6316}) and exclusive 
$B \rar K \ell^+ \ell^-,~K^\ast \ell^+ \ell^-,~\gamma  \ell^+ \ell^-,~
\pi  \ell^+ \ell^-,~\rho  \ell^+ \ell^-$
\cite{R6301,R6302,R6306,R6309,R6317,R6318} and pure leptonic 
$B \rar \ell^+ \ell^-$ decays \cite{R6305,R6319} have been studied comprehensively.

Recently, exiting results have been announced by the BaBar and Belle
Collaborations for experimental study of the $B \rar K^\ast
\ell^+ \ell^-$ decay. As far as the results for the branching ratio of the
$B \rar K^\ast \ell^+ \ell^-$ decay measured by these Collaborations are
gives as
\bea
{\cal B}(B \rar K^\ast \ell^+ \ell^-) = \left\{ \begin{array}{lc}
\left( 11.5^{+2.6}_{-2.4} \pm 0.8 \pm 0.2\right) \times
10^{-7}& \cite{R6320}~,\\ \\
\left( 0.88^{+0.33}_{-0.29} \right) \times
10^{-6}& \cite{R6321}~.\end{array} \right. \nnb
\eea 

The paper is organized as follows. In section 2, using a general form
of the effective Hamiltonian, we obtain the matrix element
in terms of the form factors of the $B \rar K^\ast$ transition. In section 3
we derive the analytical results for the polarized forward--backward
asymmetry. Last section is devoted to the numerical analysis, discussion and
conclusions.

\section{Matrix element for the $B \rar K^\ast \ell^+ \ell^-$ decay}

In this section we present the matrix element for the 
$B \rar K^\ast \ell^+ \ell^-$ decay using a general form of the
effective Hamiltonian. The $B \rar K^\ast \ell^+ \ell^-$ process is governed
by $b \rar s \ell^+ \ell^-$ transition at quark level. The effective
Hamiltonian for the $b \rar s \ell^+ \ell^-$ can be written in terms of the
twelve model independent four--Fermi interactions in the following form:
\bea
\label{e6301}
{\cal H}_{eff} \es \frac{G_F\alpha}{\sqrt{2} \pi}
 V_{ts}V_{tb}^\ast
\Bigg\{ C_{SL} \, \bar s i \sigma_{\mu\nu} \frac{q^\nu}{q^2}\, L \,b
\, \bar \ell \gamma^\mu \ell + C_{BR}\, \bar s i \sigma_{\mu\nu}
\frac{q^\nu}{q^2} \,R\, b \, \bar \ell \gamma^\mu \ell \nnb \\
\ar C_{LL}^{tot}\, \bar s_L \gamma_\mu b_L \,\bar \ell_L \gamma^\mu \ell_L +
C_{LR}^{tot} \,\bar s_L \gamma_\mu b_L \, \bar \ell_R \gamma^\mu \ell_R +
C_{RL} \,\bar s_R \gamma_\mu b_R \,\bar \ell_L \gamma^\mu \ell_L \nnb \\
\ar C_{RR} \,\bar s_R \gamma_\mu b_R \, \bar \ell_R \gamma^\mu \ell_R +
C_{LRLR} \, \bar s_L b_R \,\bar \ell_L \ell_R +
C_{RLLR} \,\bar s_R b_L \,\bar \ell_L \ell_R \nnb \\
\ar C_{LRRL} \,\bar s_L b_R \,\bar \ell_R \ell_L +
C_{RLRL} \,\bar s_R b_L \,\bar \ell_R \ell_L+
C_T\, \bar s \sigma_{\mu\nu} b \,\bar \ell \sigma^{\mu\nu}\ell \nnb \\
\ar i C_{TE}\,\epsilon^{\mu\nu\alpha\beta} \bar s \sigma_{\mu\nu} b \,
\bar \ell \sigma_{\alpha\beta} \ell  \Bigg\}~,
\eea
where $L$ and $R$ in (\ref{e6301}) are defined as
\bea
L = \frac{1-\gamma_5}{2} ~,~~~~~~ R = \frac{1+\gamma_5}{2}\nnb~,
\eea
and $C_X$ are the coefficients of the four--Fermi interactions.
Here, few words about the above Hamiltonian are in order. In principle,
${\cal O}_2$, being a member of the standard model operators, as well as 
operators of the type $\bar{s}_R b_L \bar{q}_L q_R$, where $q$ represents a
quark field, give contributions to the $b \rar s \ell^+ \ell^-$ transition
at one--loop level. The Hamiltonian given in Eq. (\ref{e6301}) should be
understood as an effective version of the most general one, where the
above--mentioned contributions are absorbed into effective Wilson coefficients
which depend on $q^2$ in general. 
The first two coefficients in Eq. (\ref{e6301}), $C_{SL}$ and $C_{BR}$, are 
the nonlocal Fermi
interactions, which correspond to $-2 m_s C_7^{eff}$ and $-2 m_b C_7^{eff}$
in the SM, respectively. The following four terms
with coefficients $C_{LL}$, $C_{LR}$, $C_{RL}$ and $C_{RR}$ are the
vector type interactions. Two of these
interactions containing $C_{LL}^{tot}$ and $C_{LR}^{tot}$ do already
exist in the SM
in the form $(C_9^{eff}-C_{10})$ and $(C_9^{eff}+C_{10})$.
Hence, representing $C_{LL}^{tot}$ and $C_{LR}^{tot}$ in the form 
\bea
C_{LL}^{tot} &=& C_9^{eff} - C_{10} + C_{LL}~, \nnb \\
C_{LR}^{tot} &=& C_9^{eff} + C_{10} + C_{LR}~, \nnb
\eea
allows us to conclude that $C_{LL}^{tot}$ and $C_{LR}^{tot}$ describe the
sum of the contributions from SM and the new physics.
The terms with
coefficients $C_{LRLR}$, $C_{RLLR}$, $C_{LRRL}$ and $C_{RLRL}$ describe
the scalar type interactions. The remaining last two terms lead by the
coefficients $C_T$ and $C_{TE}$, obviously, describe the tensor type
interactions.

The exclusive $B \rar K^\ast \ell^+ \ell^-$ decay is described in terms of
the matrix elements of the quark operators in Eq. (\ref{e6301}) over meson
states, which can be parametrized in terms of the form factors.
Obviously, the following matrix elements  
\bea
&&\lla K^\ast\vel \bar s \gamma_\mu (1 \pm \gamma_5) 
b \ver B \rra~,\nnb \\
&&\lla K^\ast \vel \bar s i\sigma_{\mu\nu} q^\nu  
(1 \pm \gamma_5) b \ver B \rra~, \nnb \\
&&\lla K^\ast \vel \bar s (1 \pm \gamma_5) b 
\ver B \rra~, \nnb \\
&&\lla K^\ast \vel \bar s \sigma_{\mu\nu} b
\ver B \rra~, \nnb
\eea
are needed for the calculation of the $B \rar K^\ast \ell^+ \ell^-$ decay. 
These matrix elements are defined as follows:
\bea
\lefteqn{
\label{e6302}
\lla K^\ast(p_{K^\ast},\varepsilon) \vel \bar s \gamma_\mu 
(1 \pm \gamma_5) b \ver B(p_B) \rra =} \nnb \\
&&- \epsilon_{\mu\nu\lambda\sigma} \varepsilon^{\ast\nu} p_{K^\ast}^\lambda q^\sigma
\frac{2 V(q^2)}{m_B+m_{K^\ast}} \pm i \varepsilon_\mu^\ast (m_B+m_{K^\ast})   
A_1(q^2) \\
&&\mp i (p_B + p_{K^\ast})_\mu (\varepsilon^\ast q)
\frac{A_2(q^2)}{m_B+m_{K^\ast}}
\mp i q_\mu \frac{2 m_{K^\ast}}{q^2} (\varepsilon^\ast q)
\left[A_3(q^2)-A_0(q^2)\right]~,  \nnb \\  \nnb \\
\lefteqn{
\label{e6303}
\lla K^\ast(p_{K^\ast},\varepsilon) \vel \bar s i \sigma_{\mu\nu} q^\nu
(1 \pm \gamma_5) b \ver B(p_B) \rra =} \nnb \\
&&4 \epsilon_{\mu\nu\lambda\sigma} 
\varepsilon^{\ast\nu} p_{K^\ast}^\lambda q^\sigma
T_1(q^2) \pm 2 i \left[ \varepsilon_\mu^\ast (m_B^2-m_{K^\ast}^2) -
(p_B + p_{K^\ast})_\mu (\varepsilon^\ast q) \right] T_2(q^2) \\
&&\pm 2 i (\varepsilon^\ast q) \left[ q_\mu -
(p_B + p_{K^\ast})_\mu \frac{q^2}{m_B^2-m_{K^\ast}^2} \right] 
T_3(q^2)~, \nnb \\  \nnb \\ 
\lefteqn{
\label{e6304}
\lla K^\ast(p_{K^\ast},\varepsilon) \vel \bar s \sigma_{\mu\nu} 
 b \ver B(p_B) \rra =} \nnb \\
&&i \epsilon_{\mu\nu\lambda\sigma}  \Bigg\{ - 2 T_1(q^2)
{\varepsilon^\ast}^\lambda (p_B + p_{K^\ast})^\sigma +
\frac{2}{q^2} (m_B^2-m_{K^\ast}^2) \Big[ T_1(q^2) - T_2(q^2) \Big] {\varepsilon^\ast}^\lambda 
q^\sigma \\
&&- \frac{4}{q^2} \Bigg[ T_1(q^2) - T_2(q^2) - \frac{q^2}{m_B^2-m_{K^\ast}^2} 
T_3(q^2) \Bigg] (\varepsilon^\ast q) p_{K^\ast}^\lambda q^\sigma \Bigg\}~. \nnb 
\eea
where $q = p_B-p_{K^\ast}$ is the momentum transfer and $\varepsilon$ is the
polarization vector of $K^\ast$ meson. 
In order to ensure finiteness of (\ref{e6302}) at $q^2=0$, 
we assume that $A_3(q^2=0) = A_0(q^2=0)$ and $T_1(q^2=0) = T_2(q^2=0)$.
The matrix element $\lla K^\ast \vel \bar s (1 \pm \gamma_5 ) b \ver B \rra$
can be calculated from Eq. (\ref{e6302}) by 
contracting both sides of Eq. (\ref{e6302}) with $q^\mu$ and using equation of
motion. Neglecting the mass of the strange quark we get
\bea
\label{e6305}
\lla K^\ast(p_{K^\ast},\varepsilon) \vel \bar s (1 \pm \gamma_5) b \ver
B(p_B) \rra =
\frac{1}{m_b} \Big[ \mp 2i m_{K^\ast} (\varepsilon^\ast q)
A_0(q^2)\Big]~.
\eea
In deriving Eq. (\ref{e6305}) we have used the relationship
\bea
2 m_{K^\ast} A_3(q^2) = (m_B+m_{K^\ast}) A_1(q^2) -
(m_B-m_{K^\ast}) A_2(q^2)~, \nnb 
\eea
which follows from the equations of motion.

Using the definition of the form factors, as given above, the amplitude of
the  $B \rar K^\ast \ell^+ \ell^-$ decay can be written as 
\bea
\lefteqn{
\label{e6306}
{\cal M}(B\rightarrow K^\ast \ell^{+}\ell^{-}) =
\frac{G \alpha}{4 \sqrt{2} \pi} V_{tb} V_{ts}^\ast }\nnb \\
&&\times \Bigg\{
\bar \ell \gamma^\mu(1-\gamma_5) \ell \, \Big[
-2 A_1 \epsilon_{\mu\nu\lambda\sigma} \varepsilon^{\ast\nu}
p_{K^\ast}^\lambda q^\sigma
 -i B_1 \varepsilon_\mu^\ast
+ i B_2 (\varepsilon^\ast q) (p_B+p_{K^\ast})_\mu
+ i B_3 (\varepsilon^\ast q) q_\mu  \Big] \nnb \\
&&+ \bar \ell \gamma^\mu(1+\gamma_5) \ell \, \Big[
-2 C_1 \epsilon_{\mu\nu\lambda\sigma} \varepsilon^{\ast\nu}
p_{K^\ast}^\lambda q^\sigma
 -i D_1 \varepsilon_\mu^\ast    
+ i D_2 (\varepsilon^\ast q) (p_B+p_{K^\ast})_\mu
+ i D_3 (\varepsilon^\ast q) q_\mu  \Big] \nnb \\
&&+\bar \ell (1-\gamma_5) \ell \Big[ i B_4 (\varepsilon^\ast
q)\Big]
+ \bar \ell (1+\gamma_5) \ell \Big[ i B_5 (\varepsilon^\ast
q)\Big]  \nnb \\
&&+4 \bar \ell \sigma^{\mu\nu}  \ell \Big( i C_T \epsilon_{\mu\nu\lambda\sigma}
\Big) \Big[ -2 T_1 {\varepsilon^\ast}^\lambda (p_B+p_{K^\ast})^\sigma +
B_6 {\varepsilon^\ast}^\lambda q^\sigma -
B_7 (\varepsilon^\ast q) {p_{K^\ast}}^\lambda q^\sigma \Big] \nnb \\
&&+16 C_{TE} \bar \ell \sigma_{\mu\nu}  \ell \Big[ -2 T_1
{\varepsilon^\ast}^\mu (p_B+p_{K^\ast})^\nu  +B_6 {\varepsilon^\ast}^\mu q^\nu -
B_7 (\varepsilon^\ast q) {p_{K^\ast}}^\mu q^\nu
\Bigg\}~,
\eea
where
\bea
\label{e6307}
A_1 &=& (C_{LL}^{tot} + C_{RL}) \frac{V}{m_B+m_{K^\ast}} -
2 (C_{BR}+C_{SL}) \frac{T_1}{q^2} ~, \nnb \\
B_1 &=& (C_{LL}^{tot} - C_{RL}) (m_B+m_{K^\ast}) A_1 - 2
(C_{BR}-C_{SL}) (m_B^2-m_{K^\ast}^2)
\frac{T_2}{q^2} ~, \nnb \\
B_2 &=& \frac{C_{LL}^{tot} - C_{RL}}{m_B+m_{K^\ast}} A_2 - 2
(C_{BR}-C_{SL})
\frac{1}{q^2}  \left[ T_2 + \frac{q^2}{m_B^2-m_{K^\ast}^2} T_3 \right]~,
\nnb \\
B_3 &=& 2 (C_{LL}^{tot} - C_{RL}) m_{K^\ast} \frac{A_3-A_0}{q^2}+
2 (C_{BR}-C_{SL}) \frac{T_3}{q^2} ~, \nnb \\
C_1 &=& A_1 ( C_{LL}^{tot} \rar C_{LR}^{tot}~,~~C_{RL} \rar
C_{RR})~,\nnb \\
D_1 &=& B_1 ( C_{LL}^{tot} \rar C_{LR}^{tot}~,~~C_{RL} \rar
C_{RR})~,\nnb \\
D_2 &=& B_2 ( C_{LL}^{tot} \rar C_{LR}^{tot}~,~~C_{RL} \rar
C_{RR})~,\nnb \\
D_3 &=& B_3 ( C_{LL}^{tot} \rar C_{LR}^{tot}~,~~C_{RL} \rar
C_{RR})~,\nnb \\
B_4 &=& - 2 ( C_{LRRL} - C_{RLRL}) \frac{ m_{K^\ast}}{m_b} A_0 ~,\nnb \\
B_5 &=& - 2 ( C_{LRLR} - C_{RLLR}) \frac{m_{K^\ast}}{m_b} A_0 ~,\nnb \\
B_6 &=& 2 (m_B^2-m_{K^\ast}^2) \frac{T_1-T_2}{q^2} ~,\nnb \\
B_7 &=& \frac{4}{q^2} \left( T_1-T_2 - 
\frac{q^2}{m_B^2-m_{K^\ast}^2} T_3 \right)~.   
\eea

From this expression of the decay amplitude, for the differential
decay width we get the following result:
\bea
\label{e6308}
\frac{d\Gamma}{d\hat{s}}(B \rar K^\ast \ell^+ \ell^-) =
\frac{G^2 \alpha^2 m_B}{2^{14} \pi^5}
\vel V_{tb}V_{ts}^\ast \ver^2 \lambda^{1/2}(1,\hat{r},\hat{s}) v
\Delta(\hat{s})~,
\eea
with
\bea
\label{e6309}
\Delta \es
\frac{2}{3 \hat{r}_{K^\ast} \hat{s}} m_B^2
\,\mbox{\rm Re}\Big[
- 6 m_B \hat{m}_\ell \hat{s} \lambda
(B_1-D_1) (B_4^\ast - B_5^\ast) \nnb \\
\ek 12 m_B^2 \hat{m}_\ell^2 \hat{s} \lambda
\Big\{ B_4 B_5^\ast + (B_3-D_2-D_3) B_1^\ast - (B_2+B_3-D_3)
D_1^\ast \Big\} \nnb \\
\ar 6 m_B^3 \hat{m}_\ell \hat{s}
(1-\hat{r}_{K^\ast}) \lambda
(B_2-D_2) (B_4^\ast - B_5^\ast) \nnb \\
\ar 12 m_B^4 \hat{m}_\ell^2 \hat{s} 
(1-\hat{r}_{K^\ast}) \lambda
(B_2-D_2) (B_3^\ast-D_3^\ast) \nnb \\
\ar 6 m_B^3 \hat{m}_\ell \lambda \hat{s}^2
(B_4-B_5) (B_3^\ast-D_3^\ast) \nnb \\
\ar 48 \hat{m}_\ell^2 \hat{r}_{K^\ast} \hat{s} \Big\{ 3 B_1 D_1^\ast +
2 m_B^4 \lambda A_1 C_1^\ast \Big\} \nnb \\
\ar 48 m_B^5 \hat{m}_\ell \hat{s}\lambda^2
(B_2+D_2) B_7^\ast C_{TE}^\ast \nnb \\
\ek 16 m_B^4 \hat{r}_{K^\ast} \hat{s} (\hat{m}_\ell^2-\hat{s}) \lambda
\Big\{ \vel A_1\ver^2 + \vel C_1\ver^2 \Big\} \nnb \\
\ek m_B^2 \hat{s} (2 \hat{m}_\ell^2-\hat{s}) \lambda
\Big\{ \vel B_4\ver^2 + \vel B_5\ver^2 \Big\} \nnb \\
\ek 48 
m_B^3 \hat{m}_\ell \hat{s} (1-\hat{r}_{K^\ast}-\hat{s}) \lambda
\Big\{(B_1+D_1) B_7^\ast C_{TE}^\ast +
2 (B_2+D_2) B_6^\ast C_{TE}^\ast \Big\} \nnb \\
\ek 6 m_B^4 \hat{m}_\ell^2 \hat{s} \lambda
\Big\{ 2 (2+2\hat{r}_{K^\ast}-\hat{s}) B_2 D_2^\ast -
\hat{s} \vel (B_3-D_3)\ver^2 \Big\} \nnb \\
\ar 96
m_B \hat{m}_\ell \hat{s} (\lambda + 12 \hat{r}_{K^\ast} \hat{s})
(B_1+D_1) B_6^\ast C_{TE}^\ast\nnb \\
\ar 8 
m_B^2 \hat{s}^2 \Big\{
v^2 \vel C_T \ver^2 + 4 (3-2 v^2) \vel C_{TE} \ver^2 \Big\}
\Big\{4 (\lambda + 12 \hat{r}_{K^\ast} \hat{s}) \vel B_6 \ver^2 \nnb \\ 
\ek 4 m_B^2 \lambda (1-\hat{r}_{K^\ast}-\hat{s}) B_6 B_7^\ast
+ m_B^4 \lambda^2 \vel B_7 \ver^2  \Big\} \nnb \\
\ek 4 m_B^2 \lambda \Big\{
\hat{m}_\ell^2 (2 - 2 \hat{r}_{K^\ast} + \hat{s} ) +
\hat{s} (1 - \hat{r}_{K^\ast} - \hat{s} ) \Big\}
(B_1 B_2^\ast + D_1 D_2^\ast) \nnb \\
\ar \hat{s} \Big\{
6 \hat{r}_{K^\ast} \hat{s} (3+v^2) + \lambda (3-v^2)
\Big\} \Big\{ \vel B_1\ver^2 + \vel D_1\ver^2 \Big\} \nnb \\
\ek 2 m_B^4 \lambda \Big\{
\hat{m}_\ell^2 [\lambda - 3 (1-\hat{r}_{K^\ast})^2] - \lambda \hat{s} \Big\}
\Big\{ \vel B_2\ver^2 + \vel D_2\ver^2 \Big\} \nnb \\
\ar 128 m_B^2 \Big\{
4 \hat{m}_\ell^2 [ 20 \hat{r}_{K^\ast} \lambda - 
12 \hat{r}_{K^\ast} (1-\hat{r}_{K^\ast})^2 - \lambda \hat{s}] \nnb \\ 
\ar \hat{s} [ 4 \hat{r}_{K^\ast} \lambda + 12 \hat{r}_{K^\ast} (1-\hat{r}_{K^\ast})^2 +
\lambda \hat{s}] \Big\}
\vel C_T\ver^2 \vel T_1\ver^2 \nnb \\
\ar 512 m_B^2 \Big\{
\hat{s} [ 4 \hat{r}_{K^\ast} \lambda + 12 \hat{r}_{K^\ast} (1-\hat{r}_{K^\ast})^2 +
\lambda \hat{s}] \nnb \\ 
\ar 8 \hat{m}_\ell^2 [ 12 \hat{r}_{K^\ast} (1-\hat{r}_{K^\ast})^2 +
\lambda (\hat{s}-8 \hat{r}_{K^\ast})] \Big\}
\vel C_{TE}\ver^2 \vel T_1\ver^2 \nnb \\
\ek 64 m_B^2 \hat{s}^2 
\Big\{ v^2 \vel C_T \ver^2
+ 4 (3 - 2 v^2) \vel C_{TE} \ver^2 \Big\}
\Big\{ 2 [ \lambda  + 12 \hat{r}_{K^\ast} (1-\hat{r}_{K^\ast})]
B_6 T_1^\ast \nnb \\ 
\ek m_B^2 \lambda (1 + 3 \hat{r}_{K^\ast} - \hat{s}) 
B_7 T_1^\ast \Big\} \nnb \\
\ar 768  m_B^3 \hat{m}_\ell \hat{r}_{K^\ast} \hat{s}
\lambda (A_1 + C_1) C_T^\ast T_1^\ast \nnb \\
\ek 192 m_B \hat{m}_\ell \hat{s}
[ \lambda  + 12 \hat{r}_{K^\ast} (1-\hat{r}_{K^\ast})]
(B_1 + D_1) C_{TE}^\ast T_1^\ast \nnb \\
\ar 192 m_B^3 \hat{m}_\ell \hat{s} \lambda
(1 + 3 \hat{r}_{K^\ast} -\hat{s}) \lambda
(B_2 + D_2) C_{TE}^\ast T_1^\ast \Big]~,
\eea
where $\hat{s}=q^2/m_B^2$, $\hat{r}=m_{K^\ast}^2/m_B^2$ and
$\lambda(a,b,c)=a^2+b^2+c^2-2ab-2ac-2bc$,
$\hat{m}_\ell=m_\ell/m_B$, $v=\sqrt{1-4\hat{m}_\ell^2/\hat{s}}$ is the
final lepton velocity.

The definition of the polarized $FB$ asymmetries will be presented in the
next section.

\section{Polarized forward--backward asymmetries of leptons}

In this section we calculate the polarized $FB$ asymmetries. For this
purpose, we define the following orthogonal unit vectors $s_i^{\pm\mu}$ in
the rest frame of $\ell^\pm$, where $i=L,N$ or $T$ correspond to
longitudinal, normal, transversal polarization directions, respectively (see
also \cite{R6301,R6308,R6310,R6314}),
\bea
\label{e6310}   
s^{-\mu}_L \es \ga 0,\vec{e}_L^{\,-}\dr =
\ga 0,\frac{\vec{p}_-}{\vel\vec{p}_- \ver}\dr~, \nnb \\
s^{-\mu}_N \es \ga 0,\vec{e}_N^{\,-}\dr = \ga 0,\frac{\vec{p}_K\times
\vec{p}_-}{\vel \vec{p}_K\times \vec{p}_- \ver}\dr~, \nnb \\
s^{-\mu}_T \es \ga 0,\vec{e}_T^{\,-}\dr = \ga 0,\vec{e}_N^{\,-}
\times \vec{e}_L^{\,-} \dr~, \nnb \\
s^{+\mu}_L \es \ga 0,\vec{e}_L^{\,+}\dr =
\ga 0,\frac{\vec{p}_+}{\vel\vec{p}_+ \ver}\dr~, \nnb \\
s^{+\mu}_N \es \ga 0,\vec{e}_N^{\,+}\dr = \ga 0,\frac{\vec{p}_K\times
\vec{p}_+}{\vel \vec{p}_K\times \vec{p}_+ \ver}\dr~, \nnb \\
s^{+\mu}_T \es \ga 0,\vec{e}_T^{\,+}\dr = \ga 0,\vec{e}_N^{\,+}
\times \vec{e}_L^{\,+}\dr~,
\eea
where $\vec{p}_\mp$ and $\vec{p}_K$ are the three--momenta of the
leptons $\ell^\mp$ and $K^\ast$ meson in the
center of mass frame (CM) of $\ell^- \,\ell^+$ system, respectively.
Transformation of unit vectors from the rest frame of the leptons to CM
frame of leptons can be accomplished by the Lorentz boost. Boosting of the
longitudinal unit vectors $s_L^{\pm\mu}$ yields
\bea
\label{e6311}
\ga s^{\mp\mu}_L \dr_{CM} \es \ga \frac{\vel\vec{p}_\mp \ver}{m_\ell}~,
\frac{E_\ell \vec{p}_\mp}{m_\ell \vel\vec{p}_\mp \ver}\dr~,
\eea
where $\vec{p}_+ = - \vec{p}_-$, $E_\ell$ and $m_\ell$ are the energy and mass
of leptons in the CM frame, respectively.
The remaining two unit vectors $s_N^{\pm\mu}$, $s_T^{\pm\mu}$ are unchanged
under Lorentz boost.

The definition of the unpolarized and normalized differential
forward--backward asymmetry is (see for example \cite{R6322})
\bea
\label{e6312}
{\cal A}_{FB} = \frac{\ds \int_{0}^{1} \frac{d^2\Gamma}{d\hat{s} dz} -
\int_{-1}^{0} \frac{d^2\Gamma}{d\hat{s} dz}}
{\ds \int_{0}^{1} \frac{d^2\Gamma}{d\hat{s} dz} +
\int_{-1}^{0} \frac{d^2\Gamma}{d\hat{s} dz}}~,
\eea
where $z=\cos\theta$ is the angle between $B$ meson and $\ell^-$ in the
center mass frame of leptons. When the spins of both leptons are taken into
account, the ${\cal A}_{FB}$ will be a function of the spins of the final
leptons and it is defined as
\bea
\label{e6313}
{\cal A}_{FB}^{ij}(\hat{s}) \es 
\Bigg(\frac{d\Gamma(\hat{s})}{d\hat{s}} \Bigg)^{-1}
\Bigg\{ \int_0^1 dz - \int_{-1}^0 dz \Bigg\}
\Bigg\{ 
\Bigg[
\frac{d^2\Gamma(\hat{s},\vec{s}^{\,-} = \vec{i},\vec{s}^{\,+} = \vec{j})}
{d\hat{s} dz} - 
\frac{d^2\Gamma(\hat{s},\vec{s}^{\,-} = \vec{i},\vec{s}^{\,+} = -\vec{j})} 
{d\hat{s} dz}
\Bigg] \nnb \\
\ek
\Bigg[
\frac{d^2\Gamma(\hat{s},\vec{s}^{\,-} = -\vec{i},\vec{s}^{\,+} = \vec{j})} 
{d\hat{s} dz} - 
\frac{d^2\Gamma(\hat{s},\vec{s}^{\,-} = -\vec{i},\vec{s}^{\,+} = -\vec{j})} 
{d\hat{s} dz}
\Bigg]
\Bigg\}~,\nnb \\ \nnb \\
\es 
{\cal A}_{FB}(\vec{s}^{\,-}=\vec{i},\vec{s}^{\,+}=\vec{j})   -
{\cal A}_{FB}(\vec{s}^{\,-}=\vec{i},\vec{s}^{\,+}=-\vec{j})  - 
{\cal A}_{FB}(\vec{s}^{\,-}=-\vec{i},\vec{s}^{\,+}=\vec{j})  \nnb \\
\ar   
{\cal A}_{FB}(\vec{s}^{\,-}=-\vec{i},\vec{s}^{\,+}=-\vec{j})~.   
\eea

Using these definitions for the double polarized $FB$ asymmetries, we get
the following results:   

\bea
\label{e6314}
{\cal A}_{FB}^{LL} \es 
\frac{2}{\hat{r}_{K^\ast}\Delta} m_B^3 \sqrt{\lambda} v \, \mbox{\rm Re}\Big[
- m_B^3 \hat{m}_\ell \lambda \Big\{ 4(B_1-D_1) B_7^\ast C_T^\ast -
(B_4+B_5) (B_2^\ast+D_2^\ast) \Big\} \nnb \\
\ar 4m_B^4 \hat{m}_\ell \lambda \Big\{
(1-\hat{r}_{K^\ast})(B_2-D_2) B_7^\ast C_T^\ast
+ \hat{s}(B_3-D_3) B_7^\ast C_T^\ast \Big\} \nnb \\
\ek \hat{m}_\ell (1-\hat{r}_{K^\ast} - \hat{s}) \Big\{     
B_1^\ast (B_4 + B_5 - 8 B_6 C_T) + D_1^\ast (B_4 + B_5 + 8 B_6 C_T)\Big\} \nnb \\
\ar 8 m_B \hat{r}_{K^\ast}\hat{s} (A_1 B_1^\ast - C_1 D_1^\ast)
+ 128 m_B^2 \hat{m}_\ell \hat{r}_{K^\ast} \hat{s}
(A_1 - C_1) B_6^\ast C_{TE}^\ast \nnb \\
\ar 2 m_B^3 \hat{s} \lambda \Big\{ (B_4 - B_5) B_7^\ast C_T^\ast +
2 (B_4 + B_5) B_7^\ast C_{TE}^\ast \Big\} \nnb \\
\ek 8 m_B^2 \hat{m}_\ell (1-\hat{r}_{K^\ast}) (1-\hat{r}_{K^\ast}-\hat{s})
(B_2 - D_2) B_6^\ast C_T^\ast \nnb \\
\ek 4 m_B (1-\hat{r}_{K^\ast}-\hat{s}) \hat{s} \Big\{
(B_4 - B_5) B_6^\ast C_T^\ast + 2 (B_4 + B_5) B_6^\ast C_{TE}^\ast \nnb \\ 
\ar 2 m_B \hat{m}_\ell (B_3 -D_3) B_6^\ast C_T^\ast 
\Big\} - 256 m_B^5 \hat{m}_\ell \hat{r}_{K^\ast} (1-\hat{r}_{K^\ast})
(A_1 - C_1) T_1^\ast C_{TE}^\ast \nnb \\
\ek 16 \hat{m}_\ell (1 - 5 \hat{r}_{K^\ast} - \hat{s})  
(B_1 - D_1) T_1^\ast C_T^\ast \nnb \\
\ar 16 m_B^2 \hat{m}_\ell (1 - \hat{r}_{K^\ast}) (1 + 3 \hat{r}_{K^\ast} - \hat{s}) 
(B_2 - D_2) T_1^\ast C_T^\ast \nnb \\
\ar 8 m_B  (1 + 3 \hat{r}_{K^\ast} - \hat{s}) \hat{s} \Big\{
2 (B_4 + B_5) T_1^\ast C_{TE}^\ast + (B_4 - B_5) T_1^\ast C_T^\ast \nnb \\
\ar 2 m_B \hat{m}_\ell (B_3 - D_3) T_1^\ast C_T^\ast \Big\}\Big]~,
\\ \nnb \\ \nnb
\label{e6315}
{\cal A}_{FB}^{LN} \es 
\frac{8}{3 \hat{r}_{K^\ast} \hat{s} \Delta} m_B^2 
\sqrt{\hat{s}}\lambda v \, \mbox{\rm Im}\Big[
- \hat{m}_\ell (B_1 D_1^\ast +
m_B^4 \lambda B_2 D_2^\ast )
+ 4 m_B^4 \hat{m}_\ell \hat{r}_{K^\ast} \sqrt{\hat{s}}
A_1 C_1^\ast \nnb \\
\ek 2 m_B \hat{s}
\Big\{ B_6 (C_T-2 C_{TE}) B_1^\ast +
B_6 (C_T+2 C_{TE}) D_1^\ast \Big\} \nnb \\
\ek m_B^5 \hat{s} \lambda 
\Big\{ B_7 (C_T-2 C_{TE}) B_2^\ast +
B_7 (C_T+2 C_{TE}) D_2^\ast \Big\} \nnb \\
\ek 16 m_B^2 \hat{m}_\ell \hat{s}
\Big( 4 \vel B_6 \ver^2 + m_B^4 \lambda \vel B_7 \ver^2 \Big)
C_T C_{TE}^\ast \nnb \\
\ar m_B^2 \hat{m}_\ell
(1 - \hat{r}_{K^\ast} - \hat{s})
(B_1 D_2^\ast + B_2 D_1^\ast) \nnb \\
\ar m_B^3
\hat{s} (1 - \hat{r}_{K^\ast} - \hat{s})
\Big\{ (B_1^\ast B_7 + 2 B_2^\ast B_6)(C_T - 2 C_{TE}) \nnb \\
\ar (D_1^\ast B_7 + 2 D_2^\ast B_6)(C_T + 
2 C_{TE}) \Big\} \nnb \\
\ek 64 m_B^2 \hat{m}_\ell \hat{s}
\Big\{ - m_B^2 (1 - \hat{r}_{K^\ast} - \hat{s}) \mbox{\rm Re}[B_6 B_7^\ast]
+ 4 \vel T_1 \ver^2 - 4 \mbox{\rm Re}[B_6 T_1^\ast] \nnb \\
\ar 2 m_B^2 (1 + 3 \hat{r}_{K^\ast} - \hat{s}) \mbox{\rm Re}[B_7 T_1^\ast]
\Big\} C_T C_{TE}^\ast \nnb \\
\ar 16 m_B^3 \hat{r}_{K^\ast} \hat{s}
\Big\{ (A_1 - C_1) C_T^\ast T_1^\ast -
2 (A_1 + C_1) C_{TE}^\ast T_1^\ast
\Big\} \nnb \\
\ar 4 m_B \hat{s}
\Big\{ B_1^\ast (C_T - 2 C_{TE}) T_1 +
D_1^\ast (C_T + 2 C_{TE}) T_1
\Big\} \nnb \\
\ek 4 m_B^3 \hat{s}
(1 + 3  \hat{r}_{K^\ast} - \hat{s})
\Big\{ B_2^\ast (C_T - 2 C_{TE}) T_1 +
D_2^\ast (C_T + 2 C_{TE}) T_1
\Big\} \Big]~, \\ \nnb \\ \nnb
\label{e6316}
{\cal A}_{FB}^{NL} \es
\frac{8}{3 \hat{r}_{K^\ast} \hat{s}\Delta} m_B^2 
\sqrt{\hat{s}}\lambda v \, \mbox{\rm Im}\Big[
- \hat{m}_\ell (B_1 D_1^\ast] +
m_B^4 \lambda B_2 D_2^\ast )
+ 4 m_B^2 \hat{m}_\ell \hat{r}_{K^\ast} \hat{s} 
A_1 C_1^\ast \nnb \\
\ar 2 m_B \hat{s}
\Big\{ B_6 (C_T+2 C_{TE}) B_1^\ast +
B_6 (C_T-2 C_{TE}) D_1^\ast \Big\} \nnb \\
\ar m_B^5 \hat{s} \lambda
\Big\{ B_7 (C_T+2 C_{TE}) B_2^\ast +
B_7 (C_T-2 C_{TE}) D_2^\ast \Big\}\nnb \\
\ar 16 m_B^2 \hat{m}_\ell \hat{s}
\Big( 4 \vel B_6 \ver^2 + m_B^4 \lambda \vel B_7 \ver^2 \Big)
C_T C_{TE}^\ast \nnb \\
\ar m_B^2 \hat{m}_\ell
(1 - \hat{r}_{K^\ast} - \hat{s})
(B_1 D_2^\ast + B_2 D_1^\ast) \nnb \\
\ek m_B^3
\hat{s} (1 - \hat{r}_{K^\ast} - \hat{s})
\Big\{ (B_1^\ast B_7 + 2 B_2^\ast B_6)(C_T + 2 C_{TE}) \nnb \\
\ar (D_1^\ast B_7 + 2 D_2^\ast B_6)(C_T - 2 C_{TE}) 
\Big\} \nnb \\
\ar 64 m_B^2 \hat{m}_\ell \hat{s}
\Big\{ - m_B^2 (1 - \hat{r}_{K^\ast} - \hat{s}) \mbox{\rm Re}[B_6 B_7^\ast]
+ 4 \vel T_1 \ver^2 - 4 \mbox{\rm Re}[B_6 T_1^\ast] \nnb \\
\ar 2 m_B^2 (1 + 3 \hat{r}_{K^\ast} - \hat{s}) \mbox{\rm Re}[B_7 T_1^\ast]
\Big\} C_T C_{TE}^\ast \nnb \\
\ar 16 m_B^3 \hat{r}_{K^\ast} \hat{s}
\Big\{ (A_1 - C_1) C_T^\ast T_1^\ast +
2 (A_1 + C_1) C_{TE}^\ast T_1^\ast
\Big\} \nnb \\
\ek 4 m_B \hat{s}
\Big\{ B_1^\ast (C_T + 2 C_{TE}) T_1 +
D_1^\ast (C_T - 2 C_{TE}) T_1
\Big\} \nnb \\
\ar 4m_B^3 \hat{s}
(1 + 3  \hat{r}_{K^\ast} - \hat{s})
\Big\{ B_2^\ast (C_T + 2 C_{TE}) T_1 +
D_2^\ast (C_T - 2 C_{TE}) T_1
\Big\} \Big]~, \\ \nnb \\ \nnb   
\label{e6317}
{\cal A}_{FB}^{LT} \es
\frac{4}{3 \hat{r}_{K^\ast} \hat{s}\Delta} m_B^2 
\sqrt{\hat{s}} \lambda \, \mbox{\rm Re} \Big[
- \hat{m}_\ell \Big\{ \vel B_1 + D_1 \ver^2 +
m_B^4 \lambda \vel B_2 + D_2 \ver^2 \Big\} \nnb \\
\ar 4  m_B^4 \hat{m}_\ell \hat{r}_{K^\ast} \hat{s}
\Big\{ \vel A_1 + C_1 \ver^2 \Big\} \nnb \\
\ek 64 m_B^2 \hat{m}_\ell \hat{s}
\vel C_{TE} \ver^2 \Big\{ 4 \vel B_6 \ver^2 + m_B^4 \lambda \vel B_7 \ver^2
- 4 m_B^2 (1 - \hat{r}_{K^\ast} - \hat{s}) B_6 B_7^\ast
\Big\} \nnb \\
\ar 2 m_B^2 \hat{m}_\ell
(1-\hat{r}_{K^\ast} - \hat{s})
(B_1+D_1) (B_2^\ast + D_2^\ast) \nnb \\
\ar 2 m_B^3 
(1-\hat{r}_{K^\ast} - \hat{s})
\Big\{ 4 \hat{m}_\ell^2 (2 B_2^\ast B_6 + B_1^\ast B_7)
(C_T + 2 C_{TE}) \nnb \\
\ek \hat{s} (2 B_2^\ast B_6 + B_1^\ast B_7) (C_T - 2 C_{TE})
\Big\} \nnb \\
\ek 4 m_B
\Big\{ 4 \hat{m}_\ell^2 \Big[ B_1^\ast B_6 (C_T + 2 C_{TE}) -
B_6 D_1^\ast (C_T - 2 C_{TE}) \Big] \nnb \\
\ek \hat{s} \Big[ B_1^\ast B_6 (C_T - 2 C_{TE}) -
B_6 D_1^\ast (C_T + 2 C_{TE}) \Big] \Big\} \nnb \\
\ek 2 m_B^5 \lambda
\Big\{ 4 \hat{m}_\ell^2 \Big[ B_2^\ast B_7 (C_T + 2 C_{TE}) -
B_7 D_2^\ast (C_T - 2 C_{TE}) \Big] \nnb \\
\ek \hat{s} \Big[ B_2^\ast B_7 (C_T - 2 C_{TE}) -
B_7 D_2^\ast (C_T + 2 C_{TE}) \Big] \Big\} \nnb \\
\ek 2 m_B^3       
(1-\hat{r}_{K^\ast} - \hat{s})
\Big\{ 4 \hat{m}_\ell^2
(2 B_6 D_2^\ast + B_7 D_1^\ast) (C_T - 2 C_{TE}) \nnb \\
\ek \hat{s} (2 B_6 D_2^\ast + B_7 D_1^\ast) (C_T + 2 C_{TE})
\Big\} \nnb \\
\ar 256 m_B^2 \hat{m}_\ell
\Big\{ 2 \hat{s} \vel C_{TE} \ver^2
\Big[ 2 B_6 T_1^\ast - 
m_B^2 (1+3 \hat{r}_{K^\ast} - \hat{s}) B_7 T_1^\ast 
\Big] \nnb \\ 
\ar 4 \vel T_1 \ver^2 \Big[ 
\hat{r}_{K^\ast} \vel C_T \ver^2 + 
(4 \hat{r}_{K^\ast} -\hat{s}) \vel C_{TE} \ver^2 \Big] \Big\} \nnb \\
\ar 32 m_B^3 \hat{r}_{K^\ast}
\Big\{ 4 \hat{m}_\ell^2
\Big[ (A_1+C_1) C_T^\ast T_1^\ast
+ 2 (A_1-C_1) C_{TE}^\ast T_1^\ast \Big] \nnb \\
\ar \hat{s} \Big[ A_1^\ast (C_T - 2 C_{TE}) T_1 +
C_1^\ast (C_T + 2 C_{TE}) T_1 \Big] 
\Big\} \nnb \\
\ar 8  m_B
\Big\{ 4 \hat{m}_\ell^2 (C_T + 2 C_{TE})-
\hat{s} (C_T - 2 C_{TE})\Big\} \Big\{B_1^\ast - m_B^2 (1+3 \hat{r}_{K^\ast} -
\hat{s}) B_2^\ast \Big\} T_1 \nnb \\
\ek 8 m_B
\Big\{ 4 \hat{m}_\ell^2 (C_T - 2 C_{TE})-
\hat{s} (C_T + 2 C_{TE})\Big\} \Big\{D_1^\ast - m_B^2 (1+3 \hat{r}_{K^\ast} -
\hat{s}) D_2^\ast \Big\} T_1 \Big]~,  \nnb \\ \\ \nnb
\label{e6318}
{\cal A}_{FB}^{TL} \es
\frac{4}{3 \hat{r}_{K^\ast} \hat{s}\Delta} m_B^2 
\sqrt{\hat{s}}\lambda \,\mbox{\rm Re}\Big[
\hat{m}_\ell \Big\{ \vel B_1 + D_1 \ver^2 +
m_B^4 \lambda \vel B_2 + D_2 \ver^2 \Big\} \nnb \\
\ek 4 m_B^4 \hat{m}_\ell \hat{r}_{K^\ast}
\Big\{ \vel A_1 + C_1 \ver^2 \Big\} \nnb \\
\ar 64 m_B^2 \hat{m}_\ell \hat{s}
\vel C_{TE} \ver^2 \Big\{ 4 \vel B_6 \ver^2 + m_B^4 \lambda \vel B_7 \ver^2
- 4 m_B^2 (1 - \hat{r}_{K^\ast} - \hat{s}) B_6 B_7^\ast
\Big\} \nnb \\
\ek 2 m_B^2 \hat{m}_\ell
(1-\hat{r}_{K^\ast} - \hat{s})
(B_1+D_1) (B_2^\ast + D_2^\ast) \nnb \\
\ar 2 m_B^3
(1-\hat{r}_{K^\ast} - \hat{s})
\Big\{ 4 \hat{m}_\ell^2 (2 B_2^\ast B_6 + B_1^\ast B_7)
(C_T - 2 C_{TE}) \nnb \\
\ek \hat{s} (2 B_2^\ast B_6 + B_1^\ast B_7) (C_T + 2 C_{TE})
\Big\} \nnb \\
\ek 4 m_B
\Big\{ 4 \hat{m}_\ell^2 \Big[ B_1^\ast B_6 (C_T - 2 C_{TE})
- B_6 D_1^\ast (C_T + 2 C_{TE}) \Big] \nnb \\
\ek \hat{s} \Big[ B_1^\ast B_6 (C_T + 2 C_{TE}) -
B_6 D_1^\ast (C_T - 2 C_{TE}) \Big] \Big\} \nnb \\
\ek 2m_B^5 \lambda
\Big\{ 4 \hat{m}_\ell^2 \Big[ B_2^\ast B_7 (C_T - 2 C_{TE}) -
B_7 D_2^\ast (C_T + 2 C_{TE}) \Big] \nnb \\
\ek \hat{s} \Big[ B_2^\ast B_7 (C_T + 2 C_{TE}) -
B_7 D_2^\ast (C_T - 2 C_{TE}) \Big] \Big\} \nnb \\
\ek 2 m_B^3      
(1-\hat{r}_{K^\ast} - \hat{s})
\Big\{ 4 \hat{m}_\ell^2
(2 B_6 D_2^\ast + B_7 D_1^\ast) (C_T + 2 C_{TE}) \nnb \\ 
\ek \hat{s} (2 B_6 D_2^\ast + B_7 D_1^\ast) (C_T - 2 C_{TE})
\Big\} \nnb \\
\ek 256 m_B^2 \hat{m}_\ell
\Big\{ 2 \hat{s} \vel C_{TE} \ver^2
\Big[ 2 B_6 T_1^\ast -
m_B^2 (1+3 \hat{r}_{K^\ast} - \hat{s}) B_7 T_1^\ast
\Big] \nnb \\ 
\ar 4 \vel T_1 \ver^2 \Big[ 
\hat{r}_{K^\ast} \vel C_T \ver^2 +
(4 \hat{r}_{K^\ast} -\hat{s}) \vel C_{TE} \ver^2 \Big] \Big\} \nnb \\
\ek 32 m_B^3 \hat{r}_{K^\ast}
\Big\{ 4 \hat{m}_\ell^2
\Big[ (A_1+C_1) C_T^\ast T_1^\ast 
- 2 (A_1-C_1) C_{TE}^\ast T_1^\ast \Big] \nnb \\
\ar \hat{s} \Big[ A_1^\ast (C_T + 2 C_{TE}) T_1 +
C_1^\ast (C_T - 2 C_{TE}) T_1 \Big] 
\Big\} \nnb \\
\ek 8 m_B
\Big\{ 4 \hat{m}_\ell^2 (C_T + 2 C_{TE})-
\hat{s} (C_T - 2 C_{TE})\Big\} \Big\{D_1^\ast - m_B^2 (1+3 \hat{r}_{K^\ast} -
\hat{s}) D_2^\ast \Big\} T_1 \nnb \\
\ar 8 m_B
\Big\{ 4 \hat{m}_\ell^2 (C_T - 2 C_{TE})-
\hat{s} (C_T + 2 C_{TE})\Big\} \Big\{B_1^\ast - m_B^2 (1+3 \hat{r}_{K^\ast} -
\hat{s}) B_2^\ast \Big\}~, T_1 \Big]  \nnb \\ \\ \nnb 
\label{e6319}
{\cal A}_{FB}^{NT} \es
\frac{2}{\hat{r}_{K^\ast} \hat{s}\Delta} m_B^2 \sqrt{\lambda}
\,\mbox{\rm Im}\Big[
m_B^3 \hat{m}_\ell \hat{s} \lambda
\Big\{ (B_4-B_5) (B_2^\ast+D_2^\ast) +
8 B_7 C_{TE} (B_1^\ast-D_1^\ast) \nnb \\ 
\ar 8 m_B^2 \hat{s} B_7^\ast C_{TE}^\ast (B_3-D_3)
\Big\} \nnb \\
\ek 2 m_B^4 \hat{m}_\ell^2 \hat{s} \lambda
(B_2+D_2) (B_3^\ast-D_3^\ast) \nnb \\
\ar 4 m_B^4 \hat{m}_\ell
(1-\hat{r}_{K^\ast}) \lambda
\Big\{ 2 m_B \hat{s} B_7^\ast C_{TE}^\ast (B_2-D_2) +
\hat{m}_\ell B_2 D_2^\ast \Big\} \nnb \\
\ar 2 m_B^2
\hat{m}_\ell^2 \hat{s} (1+ 3 \hat{r}_{K^\ast} - \hat{s})
(B_1 B_2^\ast - D_1 D_2^\ast) \nnb \\
\ar \hat{m}_\ell (1 - \hat{r}_{K^\ast} - \hat{s})
\Big\{
m_B \hat{s} \Big[ - B_1^\ast 
(B_4 - B_5 + 16 B_6 C_{TE}) \nnb \\
\ek D_1^\ast (B_4 - B_5 - 16 B_6 C_{TE})
+ 2 m_B \hat{m}_\ell (B_1+D_1) 
(B_3^\ast-D_3^\ast) \Big] \nnb \\
\ar 4 \Big[ \hat{m}_\ell B_1 D_1^\ast +
4 m_B^3 \hat{s}^2 B_6 C_{TE} (B_3^\ast-D_3^\ast) \Big]
\Big\} \nnb \\
\ek 16 m_B^3 \hat{m}_\ell \hat{s}
(1 - \hat{r}_{K^\ast}) (1 - \hat{r}_{K^\ast} - \hat{s})
(B_2 - D_2) B_6^\ast C_{TE}^\ast \nnb \\
\ar 2 m_B^2 \hat{m}_\ell^2
[\lambda +(1 - \hat{r}_{K^\ast}) (1 - \hat{r}_{K^\ast} - \hat{s})]
(B_1^\ast D_2 + B_2^\ast D_1) \nnb \\
\ar 32 m_B^3 \hat{m}_\ell \hat{s}
(1 - \hat{r}_{K^\ast}) (1 + 3 \hat{r}_{K^\ast} - \hat{s})
(B_2 -  D_2) C_{TE}^\ast T_1^\ast \nnb \\
\ek 8 m_B \hat{s}
(1 + 3 \hat{r}_{K^\ast} - \hat{s})
\Big\{ 4 \hat{m}_\ell (B_1-D_1) C_{TE}^\ast T_1^\ast
- 2 m_B \hat{s} (B_4-B_5) C_{TE}^\ast T_1^\ast \nnb \\
\ek 4 m_B^2 \hat{m}_\ell \hat{s} (B_3-D_3) C_{TE}^\ast T_1^\ast
+ m_B \hat{s} v^2 (B_4+B_5) C_T^\ast T_1^\ast
\Big\} \nnb \\
\ek 4 m_B^2  \hat{s}^2
(1 - \hat{r}_{K^\ast} - \hat{s})
\Big\{ 2 (B_4-B_5) B_6^\ast C_{TE}^\ast -
v^2 (B_4+B_5) B_6^\ast C_T^\ast
\Big\} \nnb \\
\ar 2 m_B^4 \hat{s}^2 \lambda
\Big\{ 2 (B_4-B_5) B_7^\ast C_{TE}^\ast -
v^2 (B_4+B_5) B_7^\ast C_T^\ast
\Big\} \Big]~, \\ \nnb \\ \nnb
\label{e6320}
{\cal A}_{FB}^{TN} \es
\frac{2}{\hat{r}_{K^\ast} \hat{s}\Delta} m_B^2 \sqrt{\lambda}
\,\mbox{\rm Im}\Big[
m_B^3 \hat{m}_\ell \hat{s} \lambda
\Big\{ (B_4-B_5) (B_2^\ast+D_2^\ast) +
8 B_7 C_{TE} (B_1^\ast-D_1^\ast) \nnb \\ 
\ar 8 m_B^2 \hat{s} B_7^\ast C_{TE}^\ast (B_3-D_3)
\Big\} \nnb \\
\ek 2 m_B^4 \hat{m}_\ell^2 \hat{s} \lambda
(B_2+D_2) (B_3^\ast-D_3^\ast) \nnb \\
\ar 4 
m_B^4 \hat{m}_\ell (1-\hat{r}_{K^\ast})\lambda
\Big\{ 2 m_B \hat{s} B_7^\ast C_{TE}^\ast (B_2-D_2) +
\hat{m}_\ell B_2 D_2^\ast \Big\} \nnb \\
\ar 2 m_B^2
\hat{m}_\ell^2 \hat{s} (1+ 3 \hat{r}_{K^\ast} - \hat{s})
\mbox{\rm Im} ( B_1 B_2^\ast - D_1 D_2^\ast ) \nnb \\
\ar \hat{m}_\ell (1 - \hat{r}_{K^\ast} - \hat{s})
\Big\{ 
m_B \hat{s} \Big[ B_1^\ast (B_4 - B_5 + 16 B_6 C_{TE}) \nnb \\
\ar D_1^\ast (B_4 - B_5 - 16 B_6 C_{TE})
- 2 m_B \hat{m}_\ell (B_1+D_1) 
(B_3^\ast-D_3^\ast) \Big] \nnb \\
\ar 4 \Big[ \hat{m}_\ell B_1 D_1^\ast +
4 m_B^3 \hat{s}^2 B_6 C_{TE} (B_3^\ast-D_3^\ast) \Big]
\Big\}\nnb \\
\ek 16 m_B^3 \hat{m}_\ell \hat{s} 
(1 - \hat{r}_{K^\ast}) (1 - \hat{r}_{K^\ast} - \hat{s})
(B_2 - D_2) B_6^\ast C_{TE}^\ast \nnb \\
\ar 2 m_B^2 \hat{m}_\ell^2 
[\lambda +(1 - \hat{r}_{K^\ast}) (1 - \hat{r}_{K^\ast} - \hat{s})]
(B_1^\ast D_2 + B_2^\ast D_1) \nnb \\
\ar 32 m_B^3 \hat{m}_\ell \hat{s} 
(1 - \hat{r}_{K^\ast}) (1 + 3 \hat{r}_{K^\ast} - \hat{s})
(B_2 -  D_2) C_{TE}^\ast T_1^\ast \nnb \\
\ek 8 m_B \hat{s}
(1 + 3 \hat{r}_{K^\ast} - \hat{s})
\Big\{ 4 \hat{m}_\ell (B_1-D_1) C_{TE}^\ast T_1^\ast
- 2 m_B \hat{s} (B_4-B_5) C_{TE}^\ast T_1^\ast \nnb \\
\ek 4 m_B^2 \hat{m}_\ell \hat{s} (B_3-D_3) C_{TE}^\ast T_1^\ast
+ m_B \hat{s} v^2 (B_4+B_5) C_T^\ast T_1^\ast
\Big\} \nnb \\
\ek 4 m_B^2 \hat{s}^2
(1 - \hat{r}_{K^\ast} - \hat{s})
\Big\{ 2 (B_4-B_5) B_6^\ast C_{TE}^\ast -  
v^2 (B_4+B_5) B_6^\ast C_T^\ast
\Big\} \nnb \\
\ar 2 m_B^4 \hat{s}^2 \lambda   
\Big\{ 2 (B_4-B_5) B_7^\ast C_{TE}^\ast -  
v^2 (B_4+B_5) B_7^\ast C_T^\ast
\Big\} \Big]~, \\ \nnb \\ \nnb
\label{e6321}
{\cal A}_{FB}^{NN} \es 
\frac{2}{\hat{r}_{K^\ast}\Delta} m_B^3 \sqrt{\lambda} v
\,\mbox{\rm Re}\Big[
- m_B^2 \hat{m}_\ell \lambda
\Big\{ 4 (B_1-D_1) B_7^\ast C_T^\ast +
(B_2+D_2) (B_4^\ast + B_5^\ast) \Big\} \nnb \\
\ar 4 m_B^4 \hat{m}_\ell \lambda
\Big\{ (1-\hat{r}_{K^\ast}) (B_2-D_2) B_7^\ast C_T^\ast +
\hat{s} (B_3-D_3) B_7^\ast C_T^\ast \Big\} \nnb \\
\ar 2 m_B^3 \hat{s} \lambda
\Big\{ (B_4-B_5) B_7^\ast C_T^\ast -
2 (B_4+B_5) B_7^\ast C_{TE}^\ast
\Big\} \nnb \\
\ar \hat{m}_\ell
(1-\hat{r}_{K^\ast} - \hat{s})
\Big\{ B_1^\ast (B_4+B_5+8 B_6 C_T) \nnb \\
\ar D_1^\ast (B_4+B_5-8 B_6 C_T)
\Big\} \nnb \\
\ek 8 m_B^2 \hat{m}_\ell
(1-\hat{r}_{K^\ast}) (1-\hat{r}_{K^\ast} - \hat{s})
(B_2-D_2) B_6^\ast C_T^\ast \nnb \\
\ek 4 m_B \hat{s}
(1-\hat{r}_{K^\ast} - \hat{s})
\Big\{ (B_4-B_5) B_6^\ast C_T^\ast -
2 (B_4+B_5) B_6^\ast C_{TE}^\ast \nnb \\ 
\ar 2 m_B \hat{m}_\ell (B_3-D_3) B_6^\ast C_T^\ast
\Big\} \nnb \\
\ar 16 m_B^2 \hat{m}_\ell
(1-\hat{r}_{K^\ast}) (1+3 \hat{r}_{K^\ast} - \hat{s})
(B_2-D_2) C_T^\ast T_1^\ast \nnb \\
\ar 8 m_B \hat{s}
(1+3 \hat{r}_{K^\ast} - \hat{s})
\Big\{ (B_4-B_5) C_T^\ast T_1^\ast -
2 (B_4+B_5) C_{TE}^\ast T_1^\ast \Big\} \nnb \\
\ek 16\hat{m}_\ell
(1+3 \hat{r}_{K^\ast} - \hat{s})
(B_1-D_1) C_T^\ast T_1^\ast \nnb \\
\ar 16 m_B^2 \hat{m}_\ell \hat{s}
(1+3 \hat{r}_{K^\ast} - \hat{s})
(B_3-D_3) C_T^\ast T_1^\ast \Big]~,  \\ \nnb \\ \nnb
\label{e6322}
{\cal A}_{FB}^{TT} \es
\frac{2}{\hat{r}_{K^\ast}\Delta} m_B^3 \sqrt{\lambda} v
\,\mbox{\rm Re}\Big[
m_B^2 \hat{m}_\ell \lambda
\Big\{ 4 (B_1-D_1) B_7^\ast C_T^\ast +
(B_2+D_2) (B_4^\ast + B_5^\ast)
\Big\} \nnb \\
\ek 4 m_B^4 \hat{m}_\ell
(1-\hat{r}_{K^\ast}) \lambda
(B_2-D_2) B_7^\ast C_T^\ast \nnb \\
\ek 2 m_B^3 \hat{s} \lambda
\Big\{ (B_4-B_5) B_7^\ast C_T^\ast -
2 (B_4+B_5) B_7^\ast C_{TE}^\ast \nnb \\ 
\ar 2 m_B \hat{m}_\ell (B_3-D_3) B_7^\ast C_T^\ast
\Big\} \nnb \\
\ek 2 (1-\hat{r}_{K^\ast}- \hat{s})
\Big\{\hat{m}_\ell \Big[ B_1^\ast (B_4+B_5+8 B_6 C_T) \nnb \\
\ar D_1^\ast (B_4+B_5-8 B_6 C_T) \Big] -
4 m_B \hat{s} \Big[ (B_4-B_5) B_6^\ast C_T^\ast -
2 (B_4+B_5) B_6^\ast C_{TE}^\ast \nnb \\ 
\ar 2 m_B \hat{m}_\ell (B_3-D_3) B_6^\ast C_T^\ast \Big]
\Big\} \nnb \\
\ar 8 m_B^2 \hat{m}_\ell
(1-\hat{r}_{K^\ast}) (1-\hat{r}_{K^\ast}- \hat{s})
(B_2-D_2) B_6^\ast C_T^\ast \nnb \\
\ek 16 m_B^2 \hat{m}_\ell
(1-\hat{r}_{K^\ast}) (1+3 \hat{r}_{K^\ast}- \hat{s})
(B_2-D_2) C_T^\ast T_1^\ast \nnb \\
\ek 8 m_B \hat{s}
(1+3 \hat{r}_{K^\ast}- \hat{s})
\Big\{ (B_4-B_5) C_T^\ast T_1^\ast -
2 (B_4+B_5) C_{TE}^\ast T_1^\ast
\Big\} \nnb \\
\ar 16 \hat{m}_\ell
(1+3 \hat{r}_{K^\ast}- \hat{s})
\Big\{ (B_1-D_1) C_T^\ast T_1^\ast -
m_B^2 \hat{s} (B_3-D_3) C_T^\ast T_1^\ast
\Big\} \Big]~.
\eea
In these expressions for ${\cal A}_{FB}^{ij}$, the first index in the superscript 
describes the polarization of lepton and the second index describes that of
anti--lepton.

It should be noted here that, the double--polarized $FB$ asymmetry for the 
$B \rar K \tau^+ \tau^-$ and $b \rar s \tau^+ \tau^-$ decays are calculated 
in the supersymmetric model in \cite{R6323}.

\section{Numerical analysis}    

In this section we analyze the effects of the Wilson coefficients on
the polarized $FB$ asymmetry. The input parameters we use in our numerical
calculations are: $\vel V_{tb} V_{ts}^\ast \ver =
0.0385,~m_{K^\ast}=0.892~GeV,~~m_{\tau}=1.77~GeV,~~m_{\mu}=0.106~GeV,~
~m_{b}=4.8~GeV,~~m_{B}=5.26~GeV$ and $\Gamma_B = 4.22\times 10^{-13}~GeV$. 
For the values of the Wilson coefficients we use
$C_7^{SM}=-0.313,~C_9^{SM}=4.344$ and $C_{10}^{SM}=-4.669$. It should be
noted that the above--presented value for $C_9^{SM}$ corresponds only to
short distance contributions. In addition to the short distance
contributions, it receives long distance contributions which result from 
the conversion of $\bar{c}c$ to the lepton pair. In this work we neglect 
long distance contributions. The reason for such a choice is dictated by the
fact that, in the SM the zero position of ${\cal A}_{FB}$ for the 
$B \rar K^\ast \ell^+ \ell^-$ decay is practically
independent of the form factors and is determined in terms of short distance
Wilson coefficients $C_9^{SM}$ and $C_7^{SM}$ (see \cite{R6307,R6312}) and
$s_0=3.9~GeV^2$.
For the form factors we have used the light cone QCD sum rules results
\cite{R6324,R6325}. As a result of the analysis carried out in this scheme,
the $q^2$ dependence of the form factors can be represented in terms of
three parameters as
\bea
F(q^2) = \frac{F(0)}{\ds 1-a_F \hat{s} + b_F 
    \hat{s}^2}~, \nnb
\eea
where the values of parameters $F(0)$, $a_F$ and $b_F$ for the
$B \rar K^\ast$ decay are listed in Table 1.
\begin{table}[h]
\renewcommand{\arraystretch}{1.5}
\addtolength{\arraycolsep}{3pt}
$$
\begin{array}{|l|ccc|}
\hline
& F(0) & a_F & b_F \\ \hline
A_1^{B \rar K^*} &
\phantom{-}0.34 \pm 0.05 & 0.60 & -0.023 \\
A_2^{B \rar K^*} &
\phantom{-}0.28 \pm 0.04 & 1.18 & \phantom{-}0.281\\
V^{B \rar K^*} &
 \phantom{-}0.46 \pm 0.07 & 1.55 & \phantom{-}0.575\\
T_1^{B \rar K^*} &
  \phantom{-}0.19 \pm 0.03 & 1.59 & \phantom{-}0.615\\
T_2^{B \rar K^*} &
 \phantom{-}0.19 \pm 0.03 & 0.49 & -0.241\\
T_3^{B \rar K^*} &
 \phantom{-}0.13 \pm 0.02 & 1.20 & \phantom{-}0.098\\ \hline
\end{array}
$$       
\caption{$B$ meson decay form factors in a three-parameter fit, where the
radiative corrections to the leading twist contribution and SU(3) breaking
effects are taken into account.}
\renewcommand{\arraystretch}{1}
\addtolength{\arraycolsep}{-3pt}
\end{table}
 
The new Wilson coefficients vary in the range $-\ve C_{10}\ve \le \ve C_i\ve
\le \ve C_{10}\ve$. The experimental value of the branching ratio of the $B
\rar K^\ast \ell^+ \ell^-$ decay \cite{R6320,R6321} and the bound on the
branching ratio of the $B \rar \mu^+ \mu^-$ \cite{R6326} suggest that this 
is the right order of magnitude for the vector and scalar interaction 
coefficients. It should be noted here that the experimental results lead to 
stronger restrictions on some of the Wilson coefficients, namely $-1.5 \le
C_T \le 1.5$, $-3.3 \le C_{TE} \le 2.6$, $-2 \le C_{LL}$, $C_{RL} \le 2.3$,
while the remaining coefficients vary in the range $-4 \le C_X \le 4$. Since
all existing experimental results are yet preliminary, we will vary all new
Wilson coefficients in the range $-4 \le C_X \le 4$. 

In Fig. 1(2) we present the dependence of the ${\cal A}_{FB}^{LL}$ on $q^2$
for the $B\rar K^\ast \mu^+ \mu^-$ at four fixed values of
$C_{LL}(C_{LR}):-4,-2,2,4$. From these figures we see that nonzero values
of the new Wilson coefficients shift the zero position of 
${\cal A}_{FB}^{LL}$ corresponding to the SM result. When $C_{LL}$ gets
negative (positive) values, the zero position of ${\cal A}_{FB}^{LL}$ shifts
to the left (right) in comparison to that of the zero position in the SM.

Our analysis shows that the zero position of ${\cal A}_{FB}^{LL}$ for the
$B\rar K^\ast \mu^+ \mu^-$ decay is practically independent of the existence 
of other Wilson coefficients. For this reason we do not present the
dependence of ${\cal A}_{FB}^{LL}$ on $q^2$ at fixed
values of the remaining Wilson coefficients.

Figs. 3(5) and 4(6)  depict  the dependence of ${\cal A}_{FB}^{LT}$ and 
${\cal A}_{FB}^{TL}$ on $q^2$ at four fixed values of $C_T(C_{TE})$. We
observe from these figures that the zero positions of ${\cal A}_{FB}^{LT}$
and ${\cal A}_{FB}^{TL}$ are very sensitive to the existence of tensor
interactions. More essential than is that in the SM case 
${\cal A}_{FB}^{LT}$ and ${\cal A}_{FB}^{TL}$ do not have zero values.
Therefore, if zero values for the polarized 
${\cal A}_{FB}^{LT}$ and ${\cal A}_{FB}^{TL}$ asymmetries are measured in
the experiments in future, these results are unambiguous indication of the
existence of new physics beyond the SM, more specifically, the existence of
tensor interactions.

In the case of $B\rar K^\ast \tau^+ \tau^-$ decay, the zero position for the 
double polarization asymmetries ${\cal A}_{FB}^{ij}$ is
absent for most of the new Wilson coefficients, and hence, it could be
concluded to be insensitive
to the new physics beyond the SM, or the value of ${\cal A}_{FB}^{ij}$ is
quite small, whose measurement in the experiments could practically be
impossible. For this reason we do not present the dependencies of 
${\cal A}_{FB}^{ij}$ on $q^2$ at fixed values of $C_X$ for the 
$B\rar K^\ast \tau^+ \tau^-$ decay.

As is obvious from the explicit expressions of the forward--backward
asymmetries, they depend both on $q^2$ and the new Wilson coefficients
$C_X$. As a result of this, it might be difficult to study the dependence of
the polarized forward--backward asymmetries ${\cal A}_{FB}^{ij}$ on these
parameters. However, we can eliminate the dependence of the polarized 
${\cal A}_{FB}^{ij}$ on $q^2$ by performing integration over $q^2$ in the
kinematically allowed region, so that the polarized forward--backward
asymmetry is said to be averaged. The averaged polarized forward--backward
asymmetry is defined as
\bea
\lla {\cal A}_{FB}^{ij} \rra = \frac{\ds
\int_{4 m_\ell^2}^{(m_B-m_{K^\ast})^2} {\cal A}_{FB}^{ij} 
\frac{d {\cal B}}{dq^2} dq^2 }
{\ds\int_{4 m_\ell^2}^{(m_B-m_{K^\ast})^2} \frac{d {\cal B}}{dq^2} 
dq^2 }~.\nnb
\eea 

In Fig. (7), we present the dependence of $\lla {\cal A}_{FB}^{LL} \rra$
on $C_X$ for the $B\rar K^\ast \mu^+ \mu^-$ decay. The common intersection 
point of all curves corresponds to the SM case. We observe from this figure
that, $\lla {\cal A}_{FB}^{LL} \rra$ has symmetric behavior on its 
dependence on $C_T$ and $C_{TE}$ with respect to zero position, and remains
smaller compared to the SM result. The only case for which 
$\lla {\cal A}_{FB}^{LL} \rra > \lla {\cal A}_{FB}^{LL} \rra_{SM}$ occurs
for the positive values of the vector interaction coefficients. Therefore,
if we measure in the experiments $\lla {\cal A}_{FB}^{LL} \rra > 
\lla {\cal A}_{FB}^{LL} \rra_{SM}$, it is a direct indication of new physics
beyond the SM, and this departure is to be attributed solely to the existence 
of vector type interactions.

The situation is even more conformative for the $B\rar K^\ast \tau^+ \tau^-$
case. In Figs. (9) and (10), we present the dependencies of 
$\lla {\cal A}_{FB}^{LL} \rra$ and $\lla {\cal A}_{FB}^{LT} \rra$ on the
new Wilson coefficients $C_X$. From Fig. (9) we observe that, with respect
to the zero value of the Wilson coefficients, $\lla {\cal A}_{FB}^{LL} \rra$ 
increases if $C_{RL},~C_{LRLR}$ and $C_{RR}$ increase, while it decreases
when $C_{RLLR}$ increases. 

From Fig. (9) we see that the dependence of $\lla {\cal A}_{FB}^{LT} \rra$ on 
the tensor interaction is stronger. When $C_T,~C_{TE}$ and $C_{LR}$ are
negative (positive) and vary from $-4$ to zero (from zero to $4$) 
$\lla {\cal A}_{FB}^{LT} \rra$ decrease (increase). Additionally, we
observe that with increasing values of $C_{RL}$ and $C_{RR}$, 
$\lla {\cal A}_{FB}^{LT} \rra$ increases. This figure further depicts that 
$\lla {\cal A}_{FB}^{LT} \rra$, for practical purposes, is not sensitive to
the existence of scalar interactions. On the other hand,
$\lla {\cal A}_{FB}^{NN} \rra$ and $\lla {\cal A}_{FB}^{TT} \rra$ are very
sensitive to the presence of tensor and scalar interactions (see Figs. (10)
and (11)).

It is clear from these results that several of the polarized forward--backward 
asymmetries show sizable departure from the SM results and they are
sensitive to the existence of different type of interactions. therefore,
study of these observables can be very useful in looking for new physics
beyond the SM. 

Obviously, if new physics beyond the SM exists, there hoped to be effects on 
the branching ratio besides the polarized ${\cal A}_{FB}$. Keeping in
mind that the measurement of the branching ratio is easier, one could find
it more convenient to study it for establishing new physics. But the
intriguing question is, whether there could appear situations in which the value 
of the branching ratio coincides with that of the SM result, while polarized
${\cal A}_{FB}$ does not. In order to answer this question we study the
correlation between the averaged, polarized $\lla {\cal A}_{FB} \rra$ and
branching ratio. In further analysis we vary the branching ratio of 
$B \rar K^\ast \mu^+ \mu^-~(K^\ast \tau^+ \tau^-)$ between the values 
$(1-3)\times 10^{-6}~[(1-3)\times 10^{-7}]$, which is very close to the SM
calculations. Note that, we do not take into account the experimental
results on branching ratio since they contain large errors, and it would be
better to wait for more improved experimental results. 

Our conclusion for the $B \rar K^\ast \mu^+ \mu^-$ decay, in regard to the 
above--mentioned correlated relation, is as follows (remember that, the
intersection of all curves corresponds to the SM value):

\begin{itemize}
\item for $\lla {\cal A}_{FB}^{LL} \rra$, such a region is absent for all
$C_X$,
\item for $\lla {\cal A}_{FB}^{TL} \rra$, such a region does exist for
$C_{T}$ and $C_{TE}$ (see Fig. (13)).
\end{itemize}

The situation is much more attractive for the $B \rar K^\ast \tau^+ \tau^-$
decay. In Figs. (14)--(18), we depict the dependence of the averaged,
forward--backward polarized asymmetries $\lla {\cal A}_{FB}^{LL} \rra$; 
$\lla {\cal A}_{FB}^{LT} \rra \approx - \lla {\cal A}_{FB}^{TL} \rra$; 
$\lla {\cal A}_{FB}^{NT} \rra \approx \lla {\cal A}_{FB}^{TN} \rra$; 
$\lla {\cal A}_{FB}^{NN} \rra$ and $\lla {\cal A}_{FB}^{TT} \rra$, on 
branching ratio. It follows from these figures that, indeed, there exist
certain regions of the new Wilson coefficients for which, mere study of the
polarized ${\cal A}_{FB}$ can give promising information about new
physics beyond the SM.

In summary, in this work we present the analysis for the forward--backward
asymmetries when both leptons are polarized, using a general, model
independent form of the effective Hamiltonian. Our work verifies that the
study of the zero position of $\lla {\cal A}_{FB}^{LL} \rra$ can give
unambiguous conformation of the new physics beyond the SM, since when new
physics effects are taken into account, the results are shifted with respect to 
their zero positions in the SM. Moreover, we find that the polarized 
${\cal A}_{FB}$ is quite sensitive to the existence of the tensor and vector
interactions. Finally we obtain that there exist certain regions of the new
Wilson coefficients for which, only study of the polarized forward--backward
asymmetry gives invaluable information in establishing new physics beyond
the SM.

\newpage

\section*{Figure captions}
{\bf Fig. (1)} The dependence of the double--lepton polarization asymmetry
${\cal A}_{FB}^{LL}$ on $q^2$ at four fixed
values of $C_{LL}$, for the $B \rar K^\ast \mu^+ \mu^-$ decay.\\ \\
{\bf Fig. (2)} The same as in Fig. (1), but at four fixed 
values of $C_{LR}$.\\ \\
{\bf Fig. (3)} The dependence of the double--lepton polarization asymmetry
${\cal A}_{FB}^{LT}$ on $q^2$ at four fixed 
values of $C_{T}$, for the $B \rar K^\ast \mu^+ \mu^-$ decay.\\ \\
{\bf Fig. (4)} The same as in Fig. (3), but for  ${\cal A}_{FB}^{TL}$.\\ \\
{\bf Fig. (5)} The same as in Fig. (3), but at four fixed 
values of $C_{TE}$.\\ \\
{\bf Fig. (6)} The same as in Fig. (4), but at four fixed 
values of $C_{TE}$.\\ \\
{\bf Fig. (7)} The dependence of the averaged forward--backward
double--lepton polarization asymmetry $\lla {\cal A}_{FB}^{LL} \rra$
on the new Wilson coefficients $C_X$, for the $B \rar K^\ast \mu^+ \mu^-$
decay.\\ \\
{\bf Fig. (8)} The same as in Fig. (7), but for the
$B \rar K^\ast \tau^+ \tau^-$ decay.\\ \\
{\bf Fig. (9)} The same as in Fig. (8), but for the averaged 
forward--backward double--lepton polarization asymmetry
$\lla {\cal A}_{FB}^{LT} \rra$.\\ \\
{\bf Fig. (10)} The same as in Fig. (8), but for the averaged 
forward--backward double--lepton polarization asymmetry
$\lla {\cal A}_{FB}^{TL} \rra$.\\ \\
{\bf Fig. (11)} The same as in Fig. (8), but for the averaged 
forward--backward double--lepton polarization asymmetry
$\lla {\cal A}_{FB}^{NN} \rra$.\\ \\
{\bf Fig. (12)} The same as in Fig. (8), but for the averaged 
forward--backward double--lepton polarization asymmetry
$\lla {\cal A}_{FB}^{TT} \rra$.\\ \\
{\bf Fig. (13)} Parametric plot of the correlation between the averaged 
forward--backward double--lepton polarization asymmetry
$\lla {\cal A}_{FB}^{LT} \rra$ and the branching ratio for the
$B \rar K^\ast \mu^+ \mu^-$ decay.\\ \\
{\bf Fig. (14)}  Parametric plot of the correlation between the averaged
forward--backward double--lepton polarization asymmetry
$\lla {\cal A}_{FB}^{LL} \rra$ and the branching ratio for the
$B \rar K^\ast \tau^+ \tau^-$ decay.\\ \\
{\bf Fig. (15)} The same as in Fig. (14), but for the 
the correlation between the averaged
forward--backward double--lepton polarization asymmetry
$\lla {\cal A}_{FB}^{LT} \rra$ and the branching ratio.\\ \\
{\bf Fig. (16)} The same as in Fig. (15), but for the 
the correlation between the averaged
forward--backward double--lepton polarization asymmetry
$\lla {\cal A}_{FB}^{NT} \rra$ and the branching ratio.\\ \\
{\bf Fig. (17)} The same as in Fig. (16), but for the 
the correlation between the averaged
forward--backward double--lepton polarization asymmetry
$\lla {\cal A}_{FB}^{NN} \rra$ and the branching ratio.\\ \\
{\bf Fig. (18)} The same as in Fig. (17), but for the 
the correlation between the averaged
forward--backward double--lepton polarization asymmetry
$\lla {\cal A}_{FB}^{TT} \rra$ and the branching ratio.\\ \\

\newpage

\begin{figure}
\vskip 1.5 cm
    \includegraphics{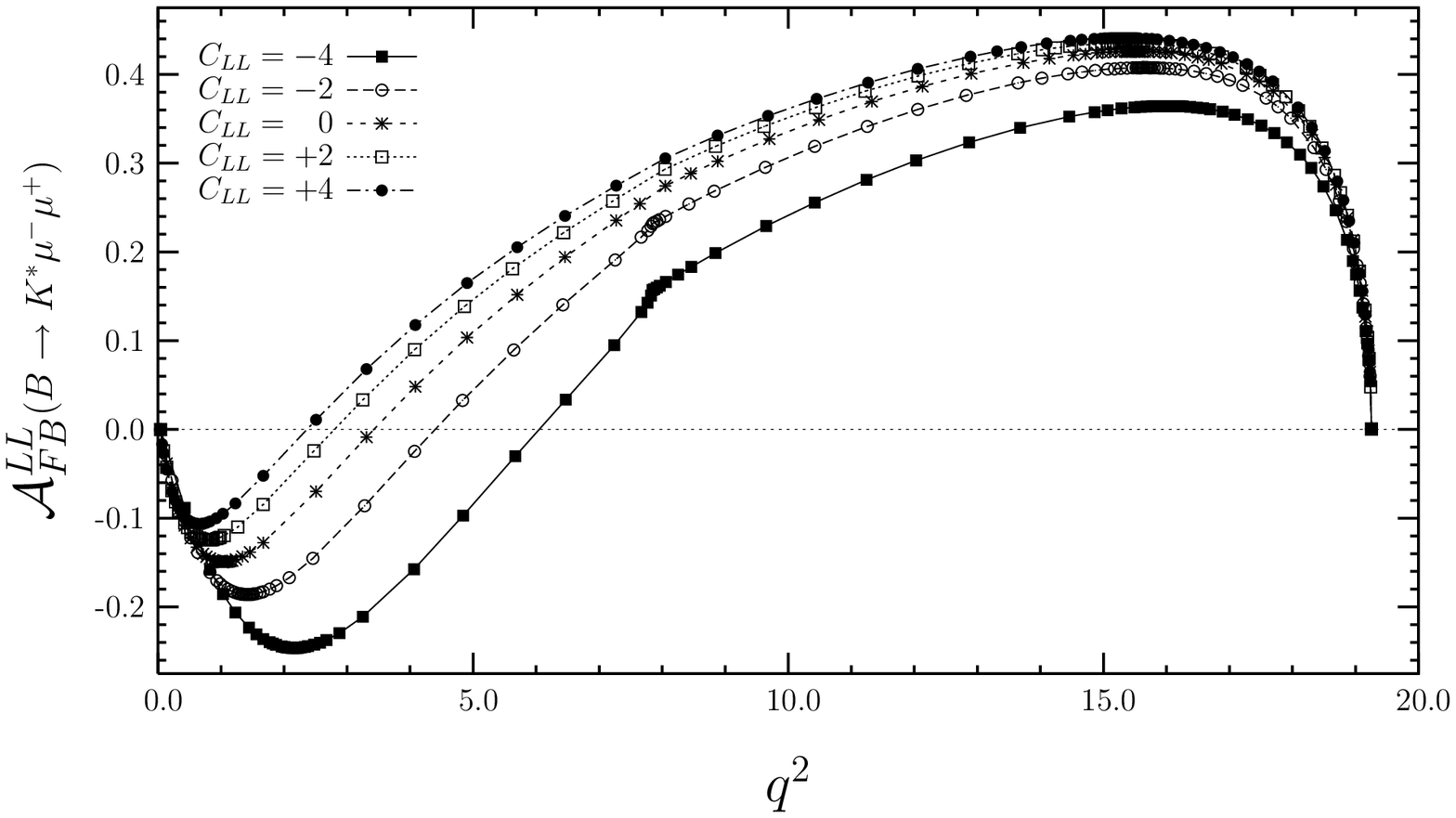}
\vskip 7.8cm
\caption{}
\end{figure}

\begin{figure}
\vskip 2.5 cm
    \includegraphics{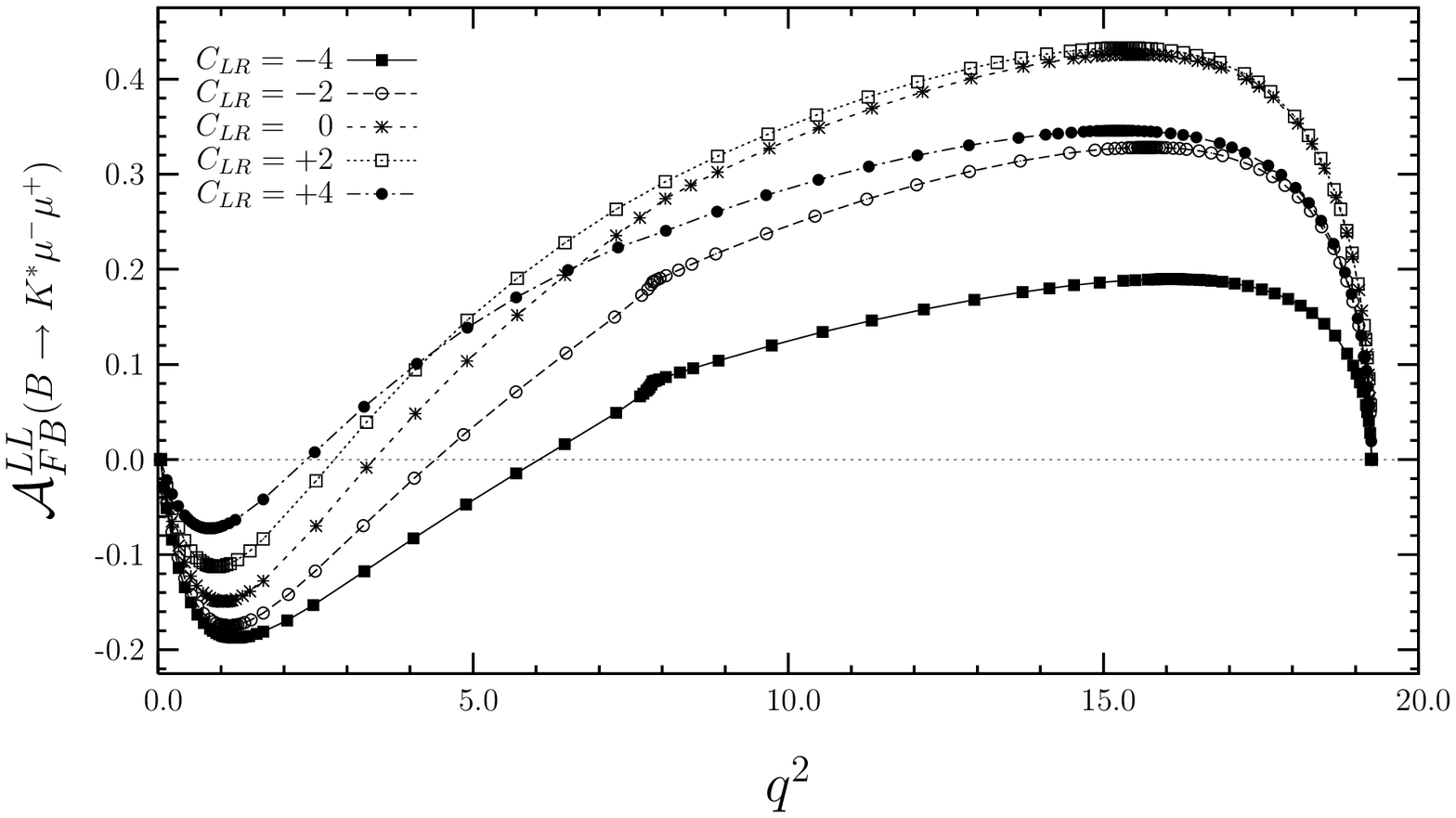}
\vskip 7.8 cm
\caption{}
\end{figure}

\begin{figure}
\vskip 1.5 cm
    \includegraphics{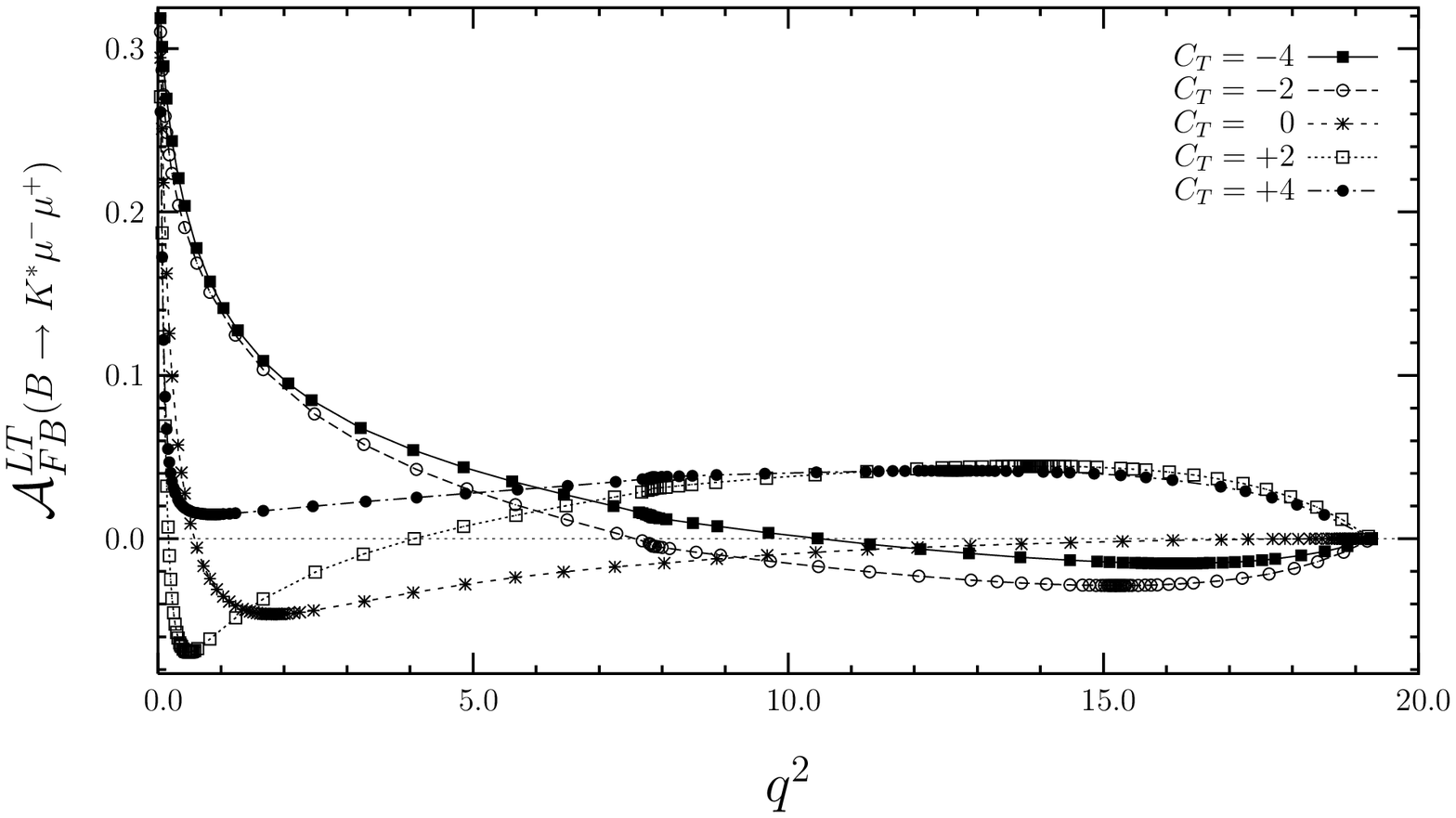}
\vskip 7.8cm
\caption{}
\end{figure}

\begin{figure}
\vskip 2.5 cm
    \includegraphics{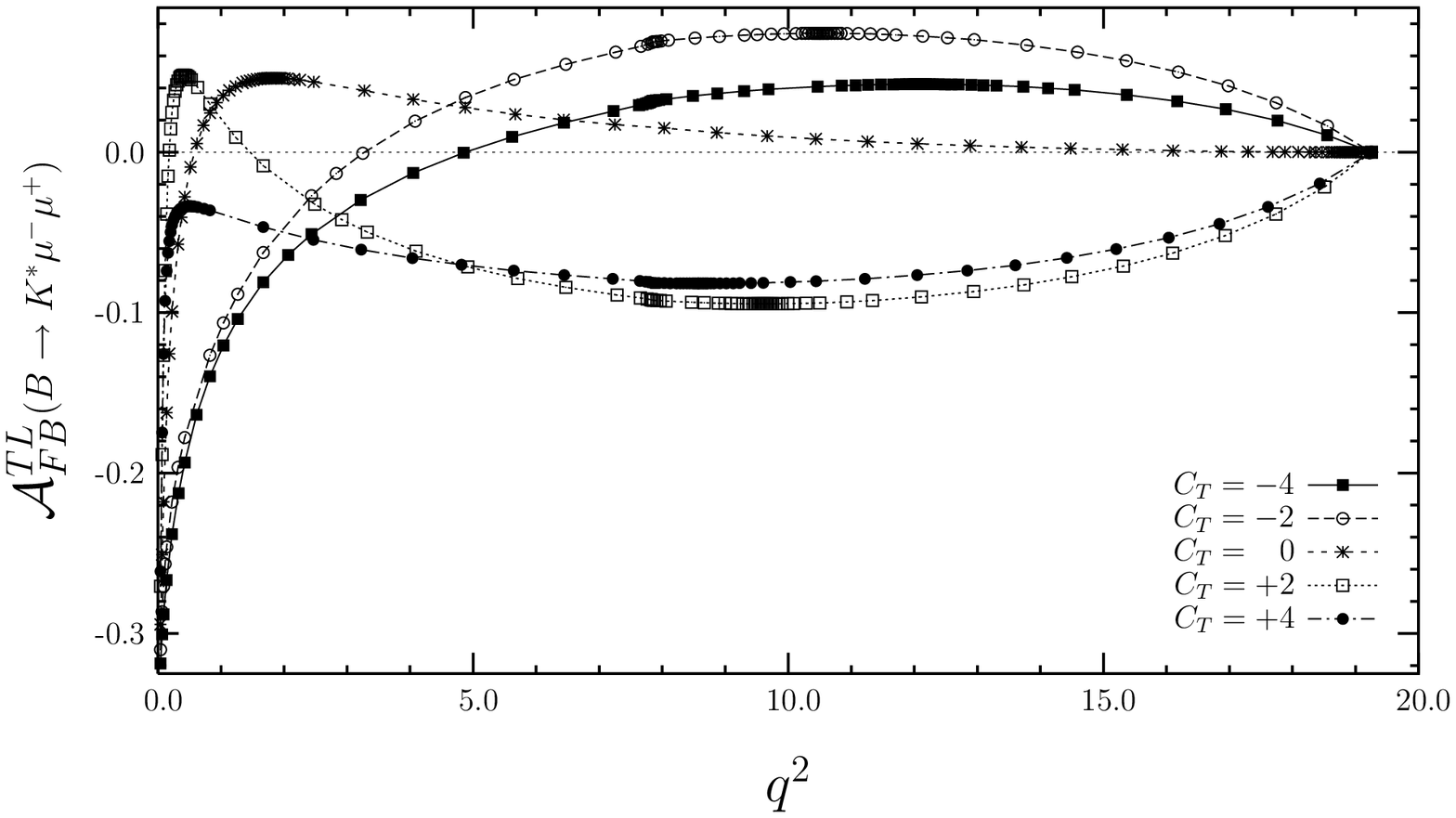}
\vskip 7.8 cm
\caption{}
\end{figure}

\begin{figure}
\vskip 2.5 cm
    \includegraphics{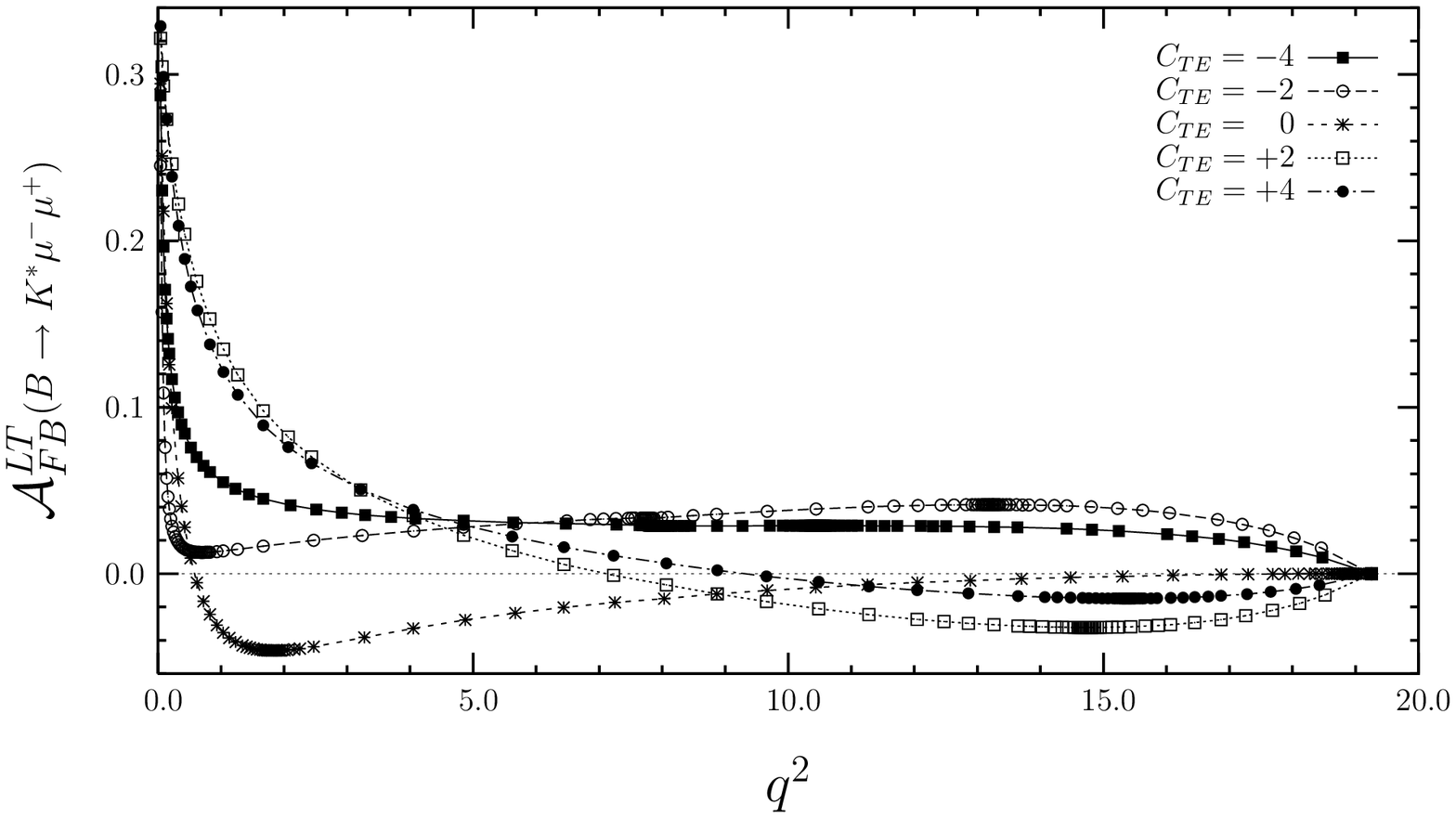}
\vskip 7.8 cm
\caption{}
\end{figure}

\begin{figure}
\vskip 1.5 cm
    \includegraphics{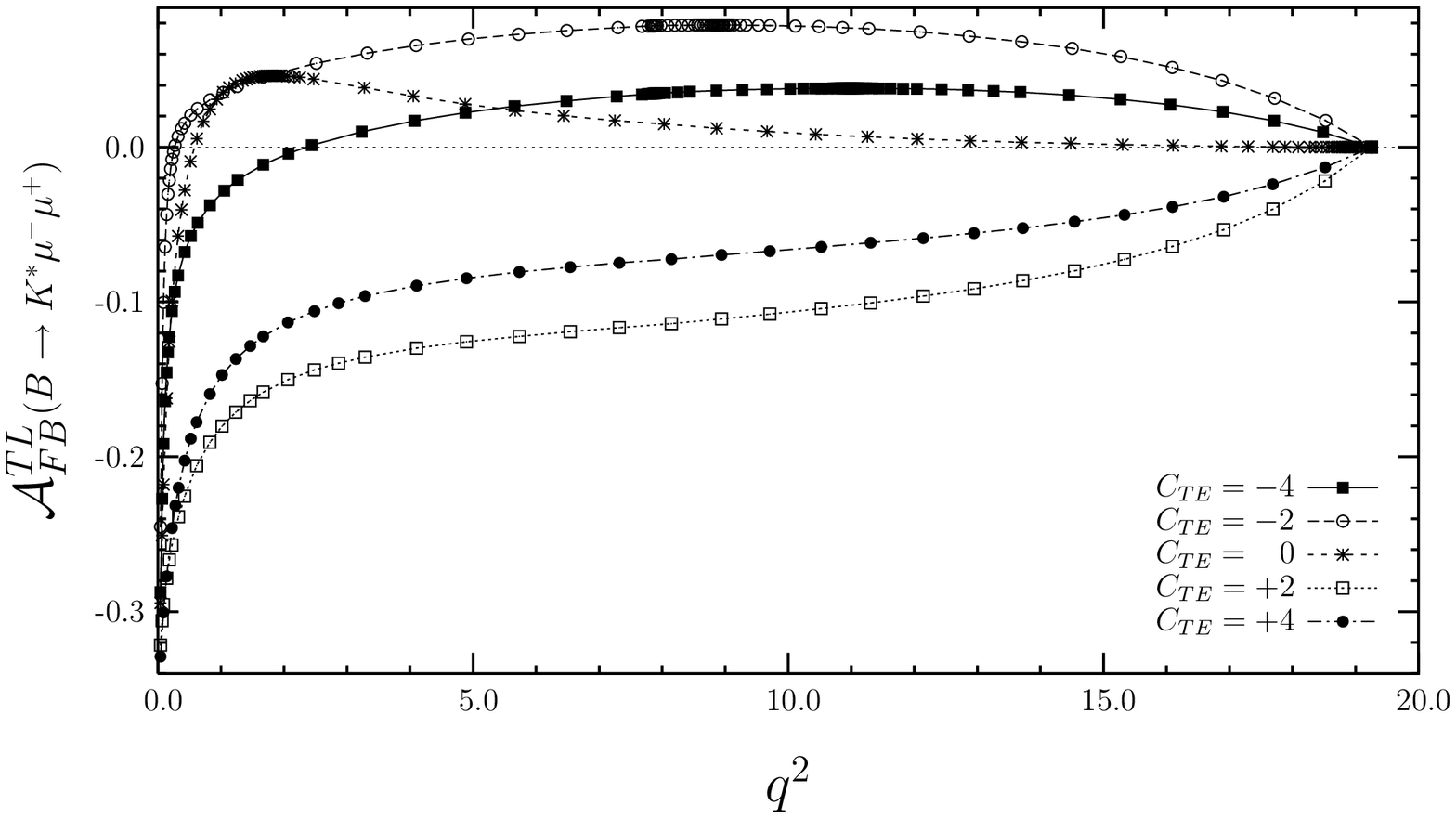}
\vskip 7.8cm
\caption{}
\end{figure}

\begin{figure}
\vskip 2.5 cm
    \includegraphics{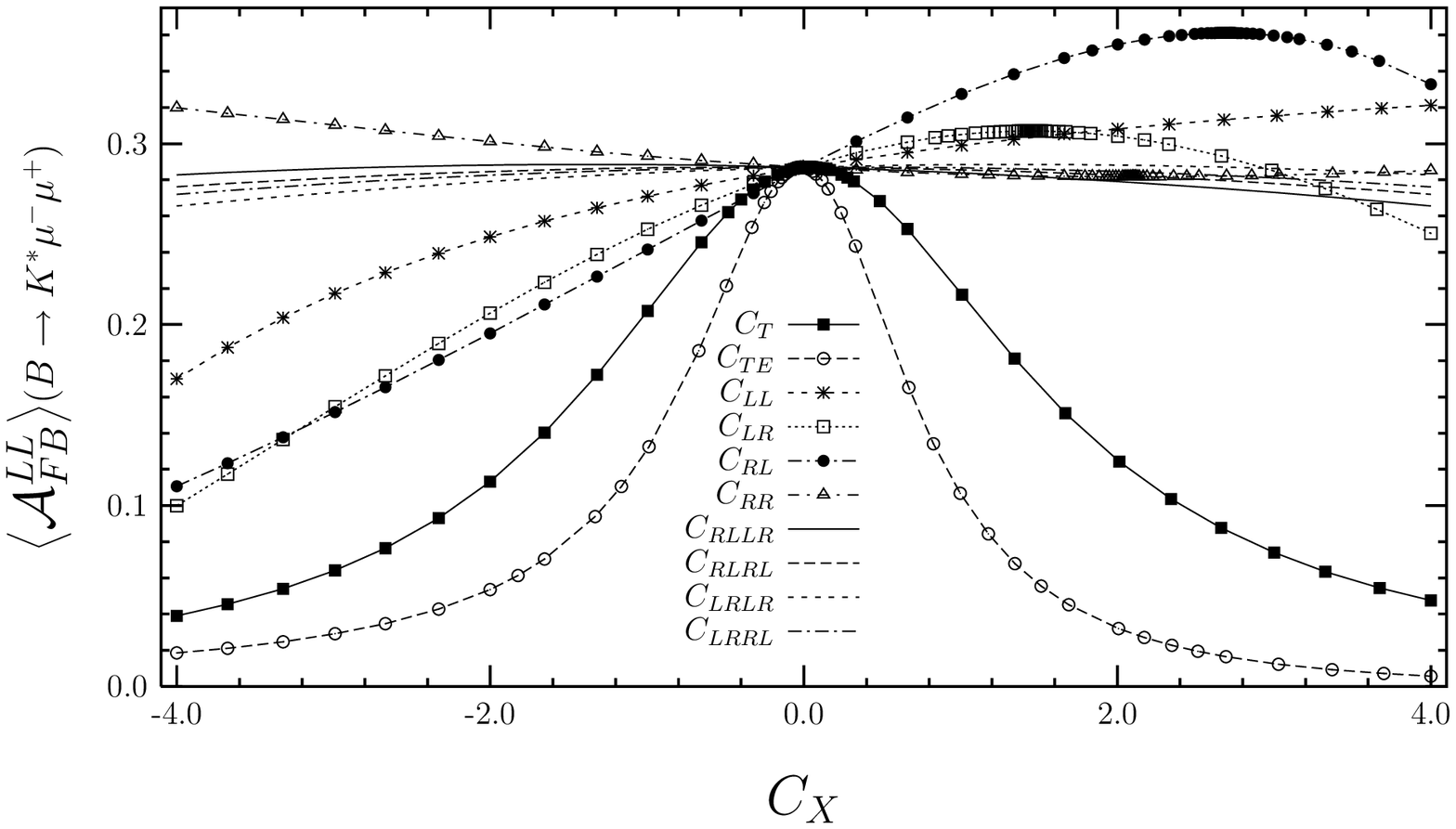}
\vskip 7.8 cm
\caption{}
\end{figure}

\begin{figure}
\vskip 1.5 cm
    \includegraphics{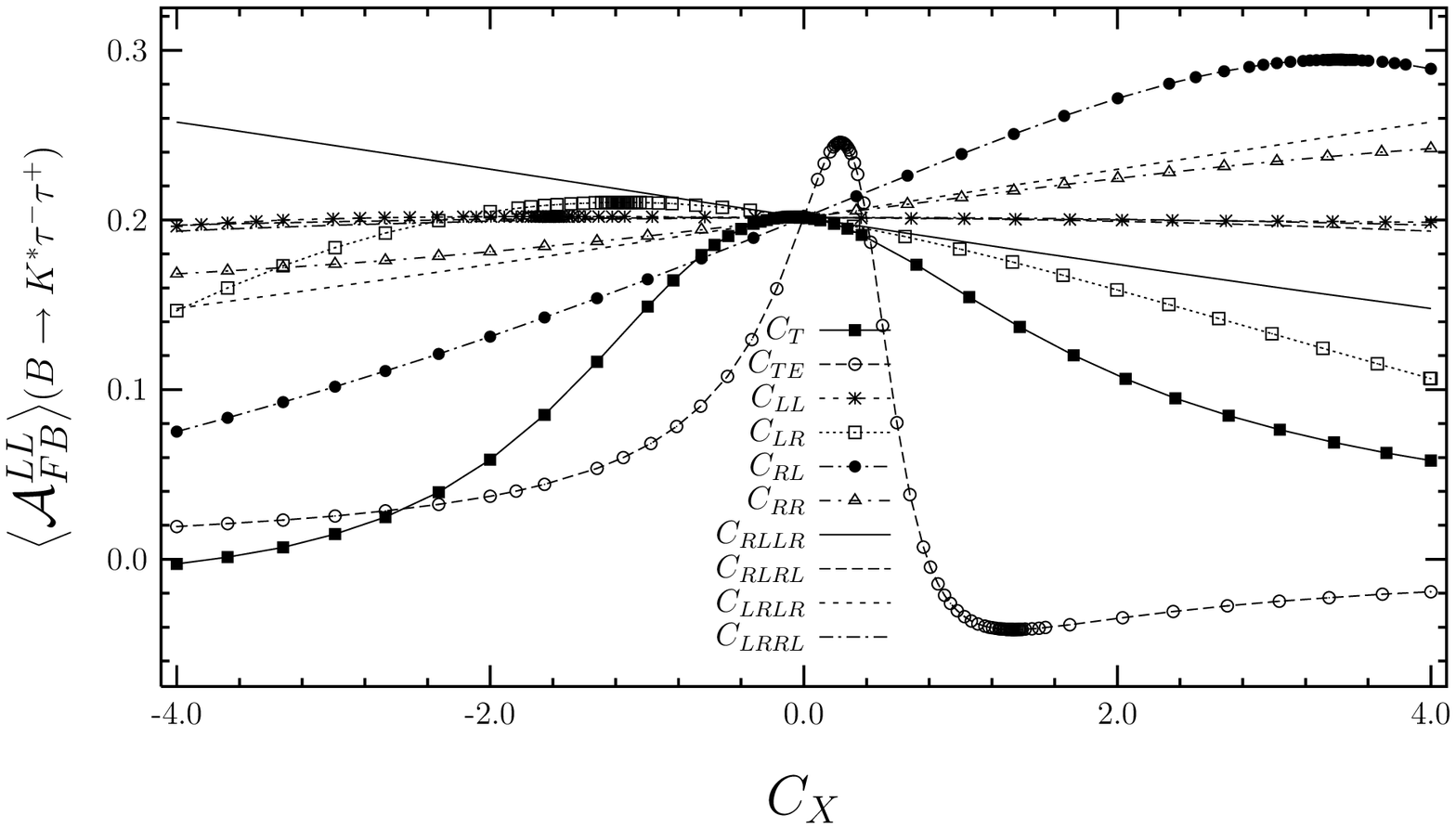}
\vskip 7.8cm
\caption{}
\end{figure}

\begin{figure}
\vskip 2.5 cm
    \includegraphics{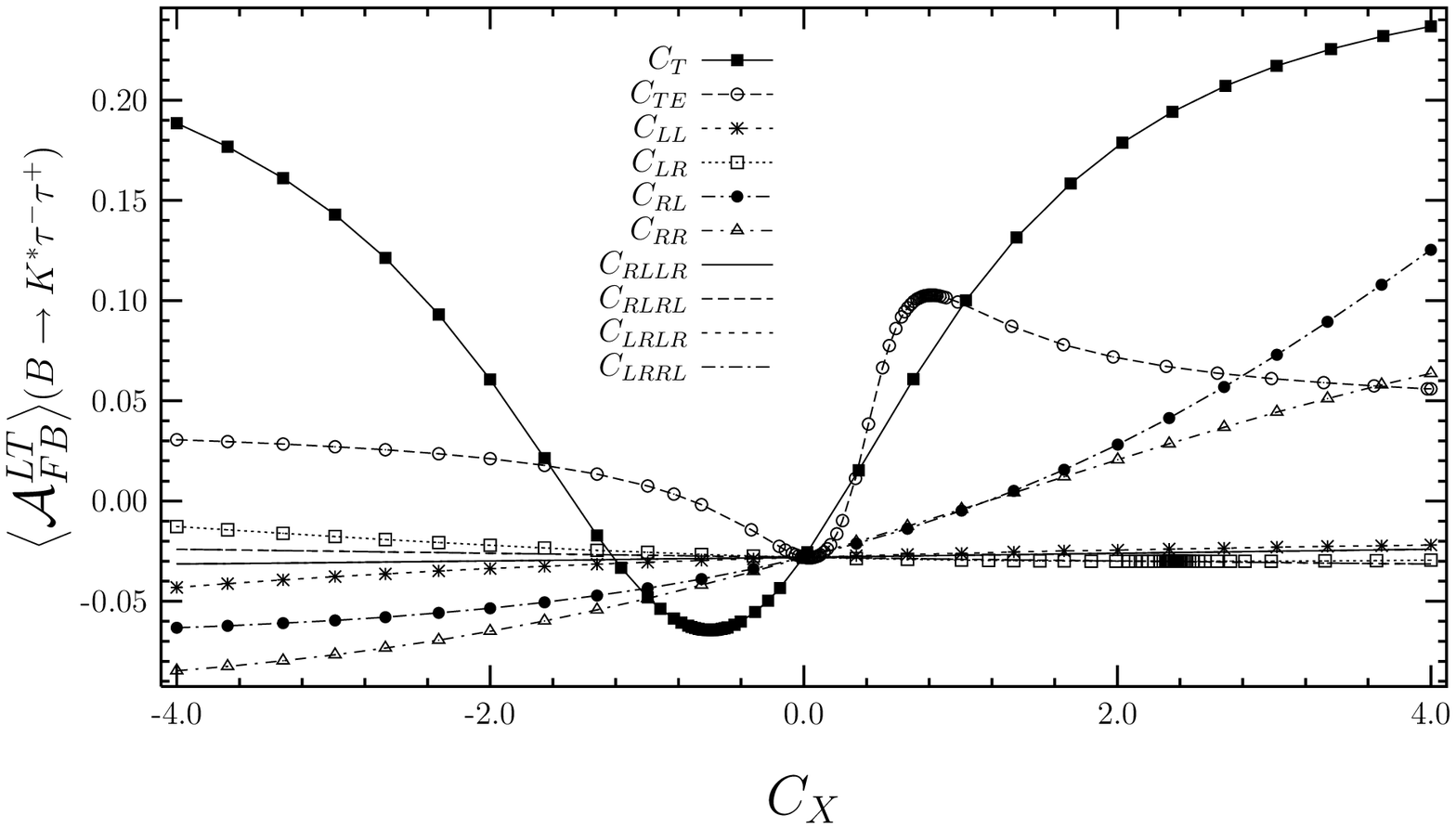}
\vskip 7.8 cm
\caption{}
\end{figure}

\begin{figure}
\vskip 1.5 cm
    \includegraphics{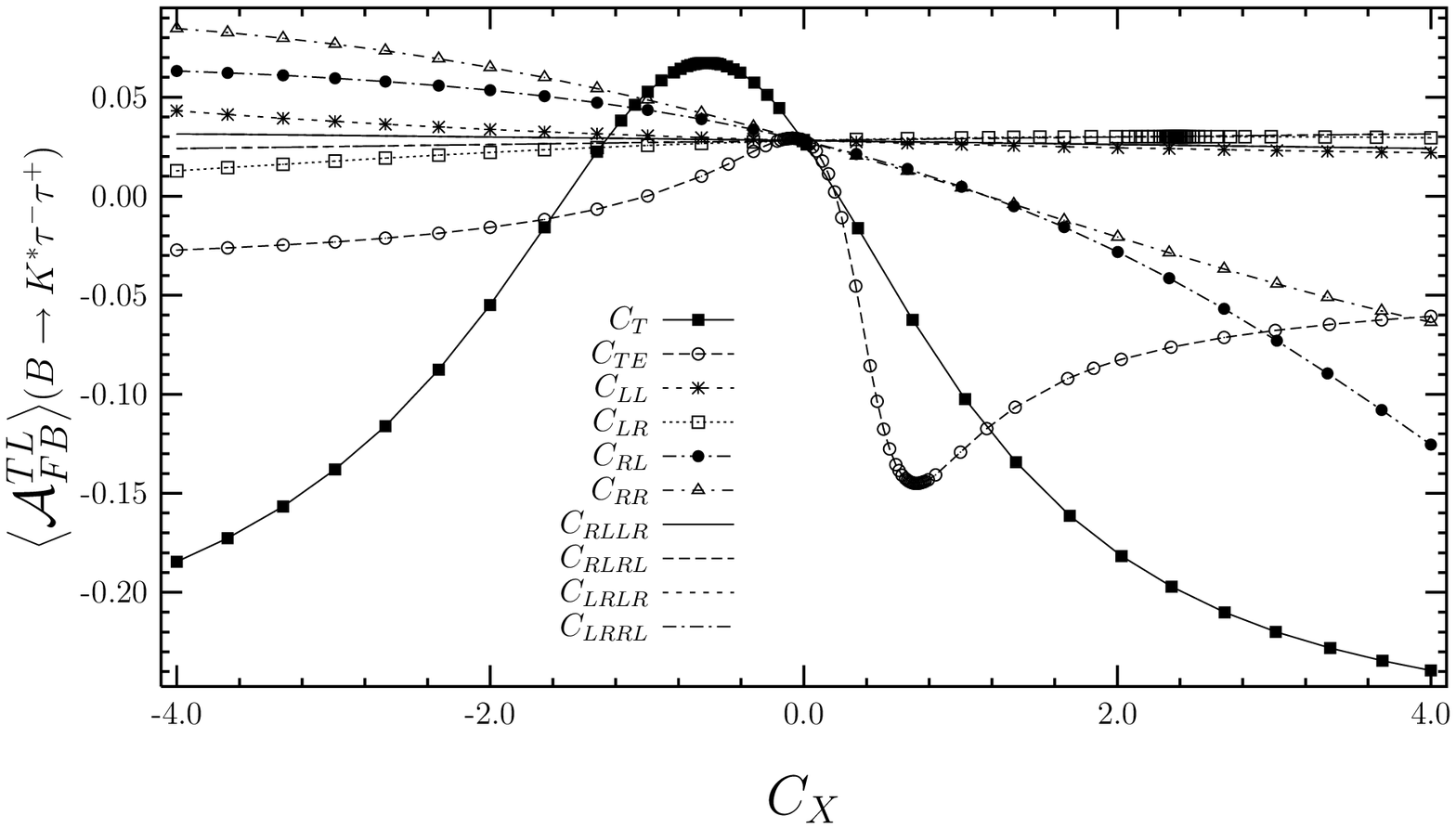}
\vskip 7.8cm
\caption{}
\end{figure}

\begin{figure}
\vskip 2.5 cm
    \includegraphics{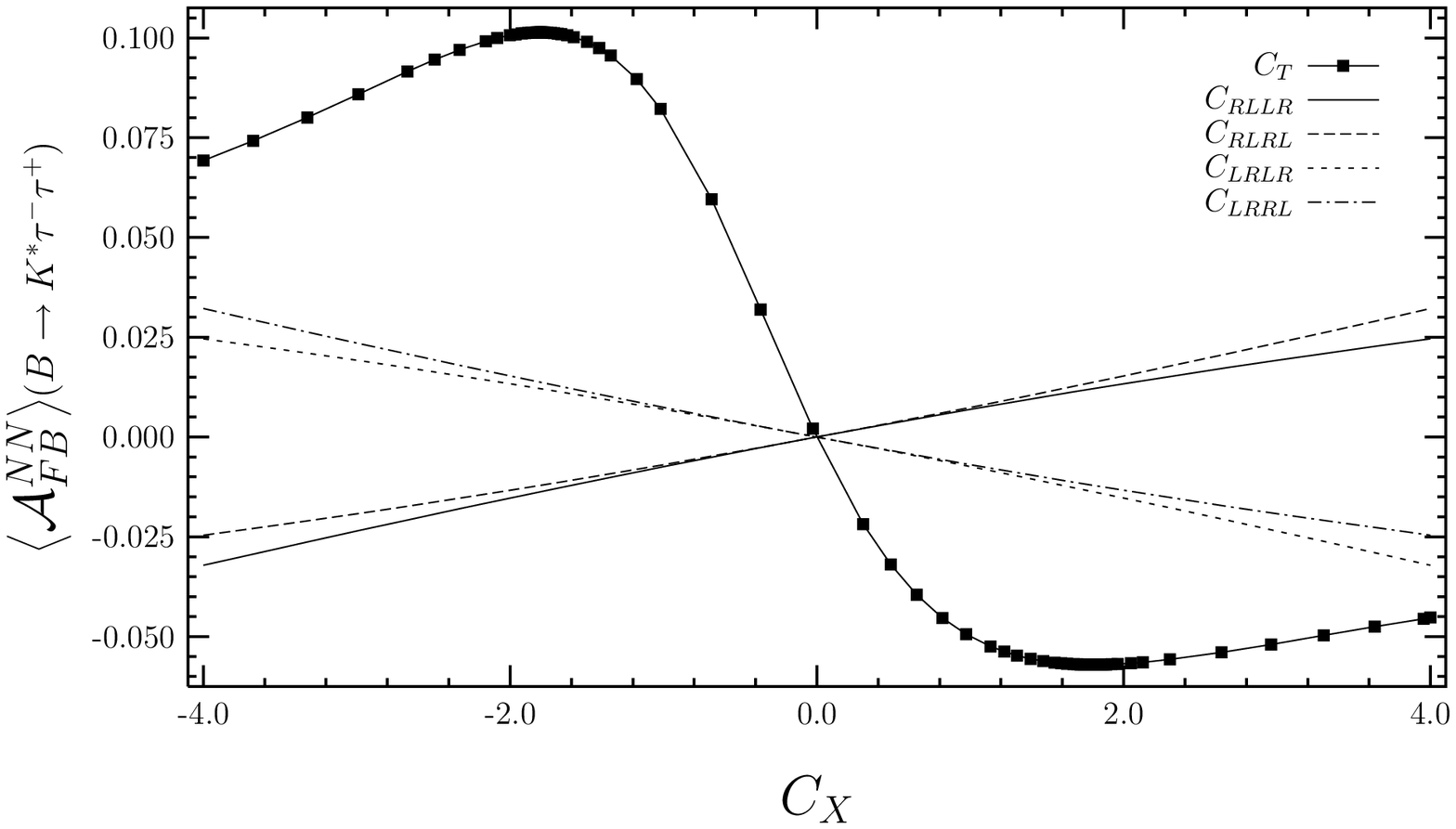}
\vskip 7.8 cm
\caption{}
\end{figure}

\begin{figure}
\vskip 1.5 cm
    \includegraphics{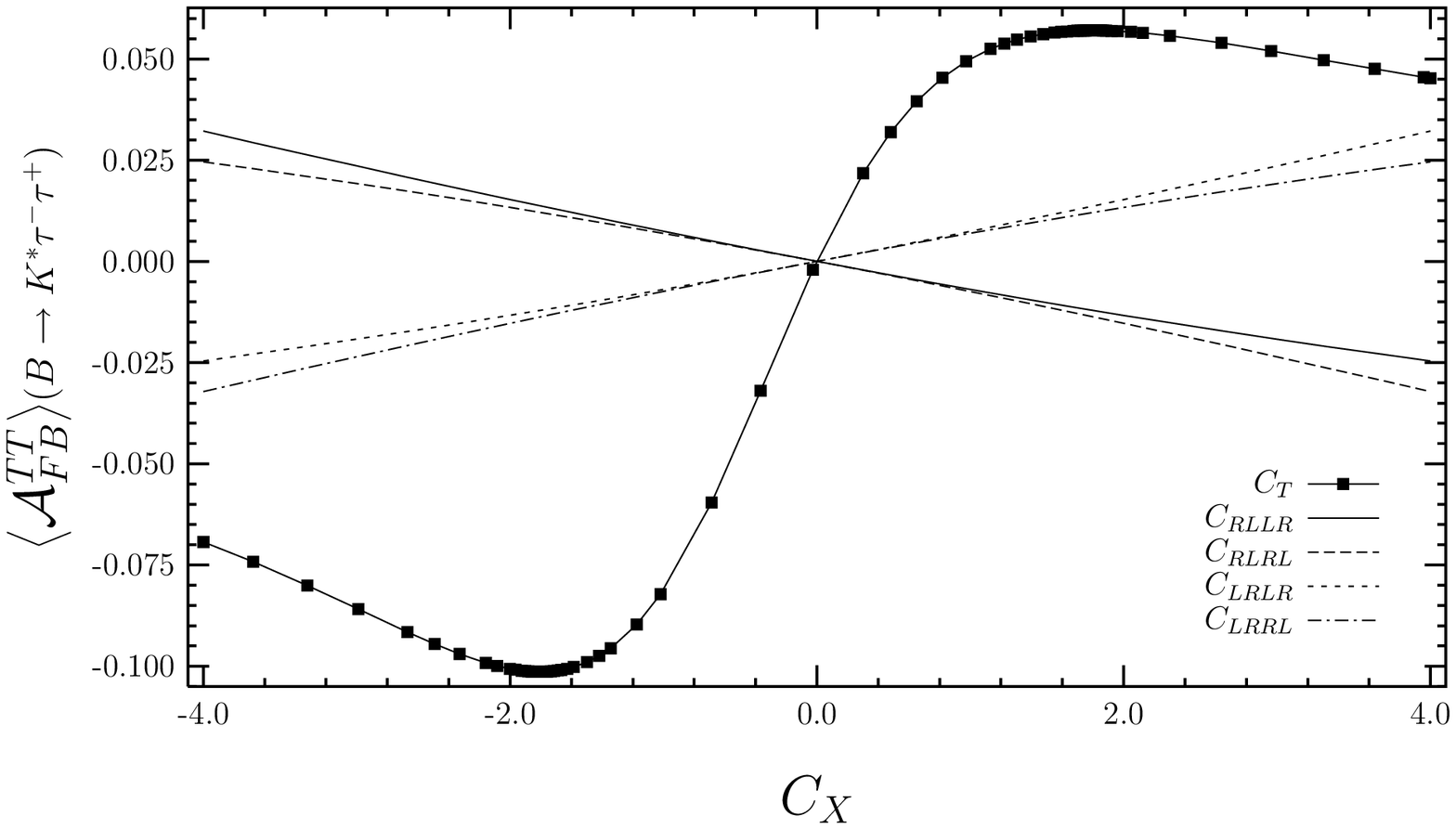}
\vskip 7.8cm
\caption{}
\end{figure}

\begin{figure}
\vskip 1.5 cm
    \includegraphics{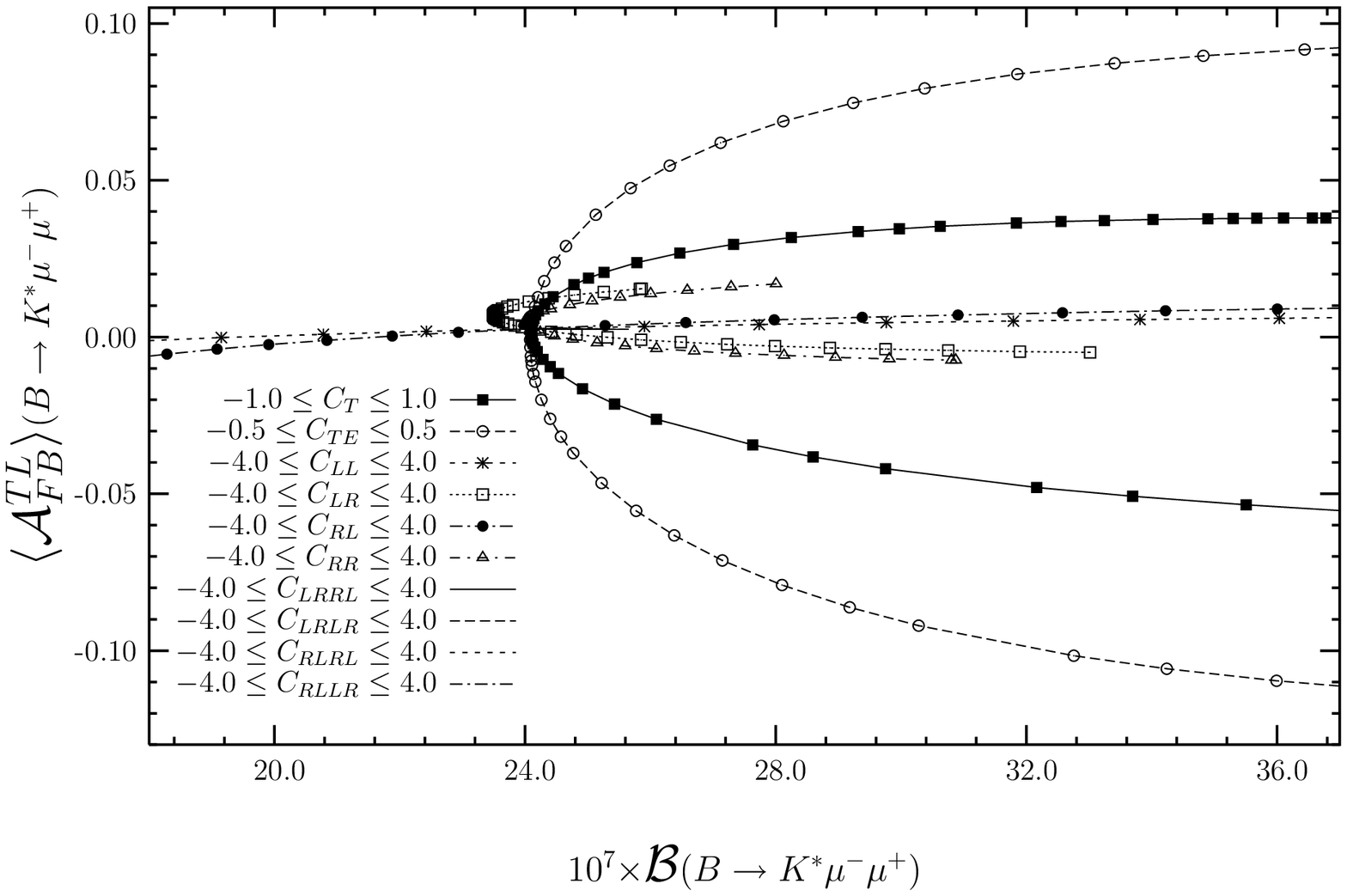}
\vskip 8.8 cm
\caption{}
\end{figure}

\begin{figure}
\vskip 2.5 cm
    \includegraphics{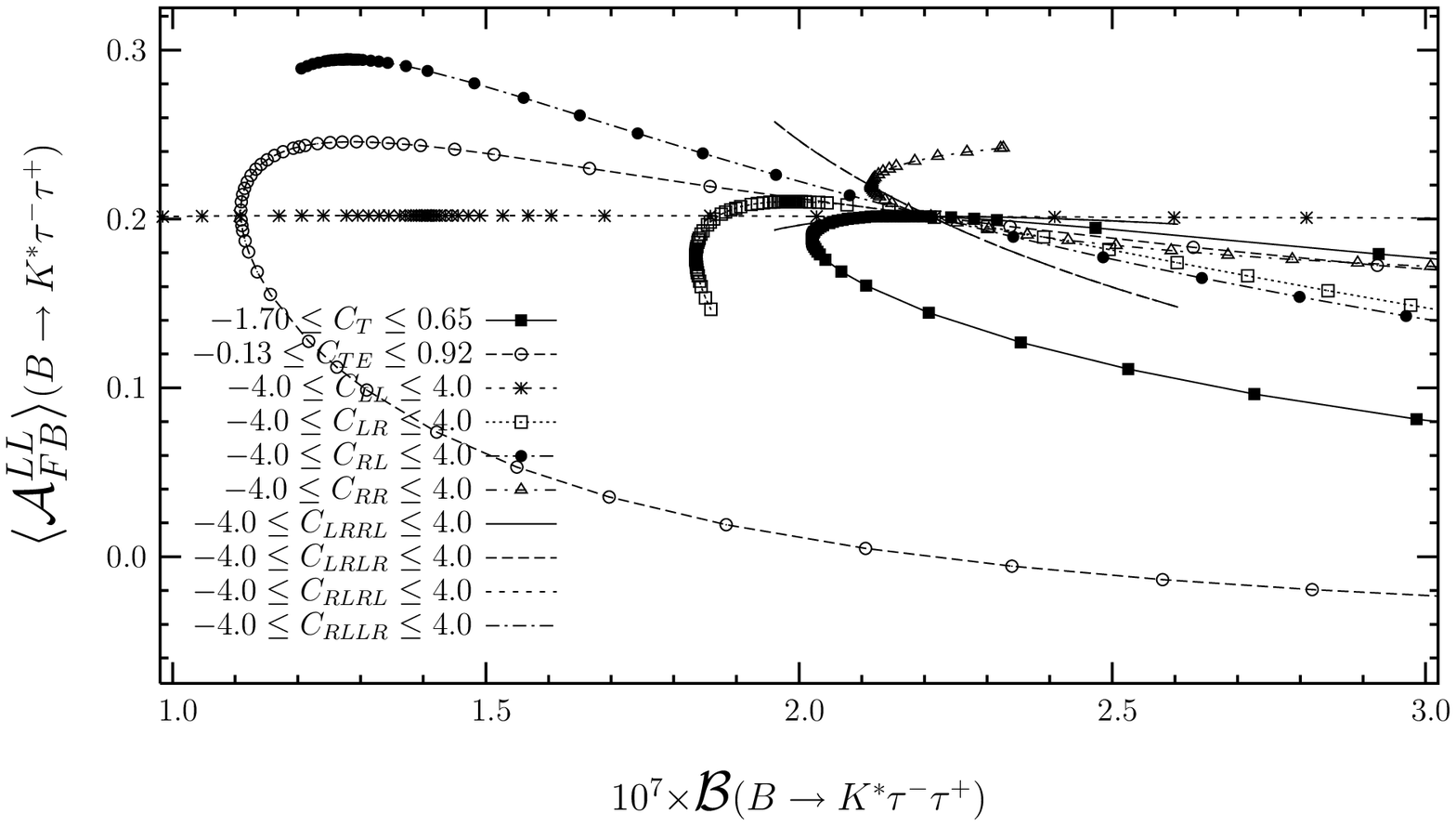}
\vskip 7.8cm
\caption{}
\end{figure}

\begin{figure}
\vskip 2.5 cm
    \includegraphics{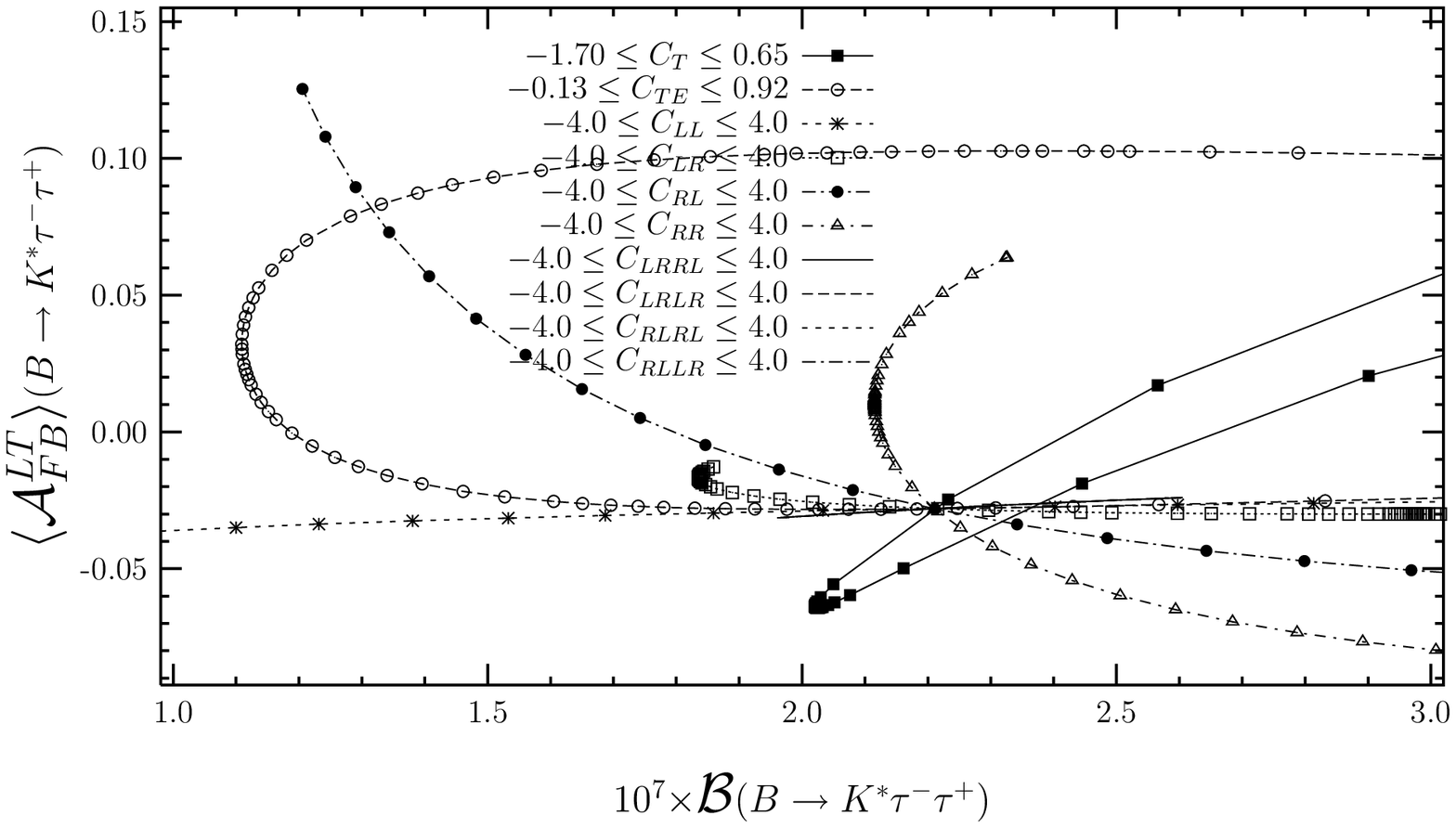}
\vskip 7.8 cm
\caption{}
\end{figure}

\begin{figure}
\vskip 1.5 cm
    \includegraphics{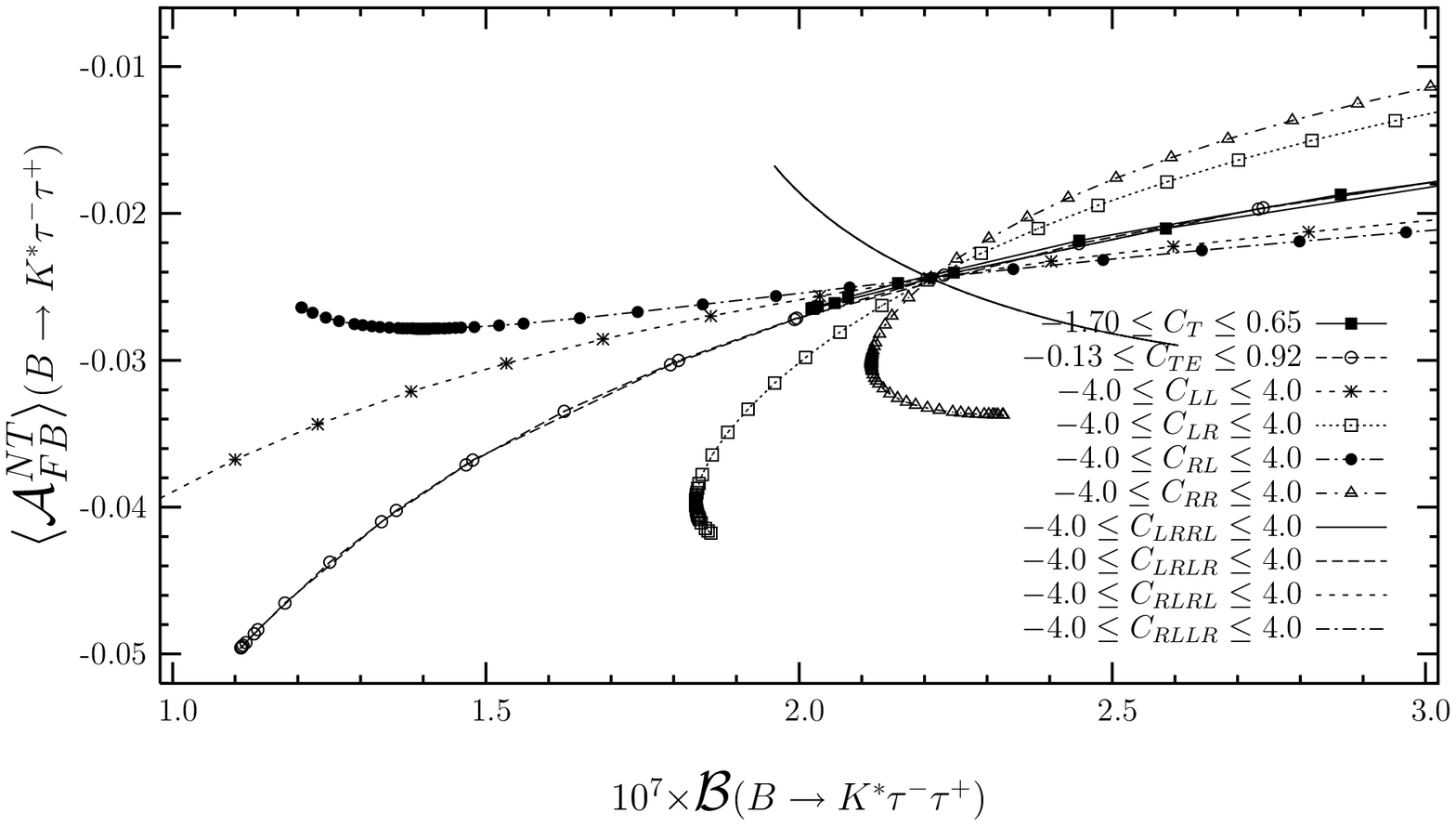}
\vskip 7.8cm
\caption{}
\end{figure}

\begin{figure}
\vskip 2.5 cm
    \includegraphics{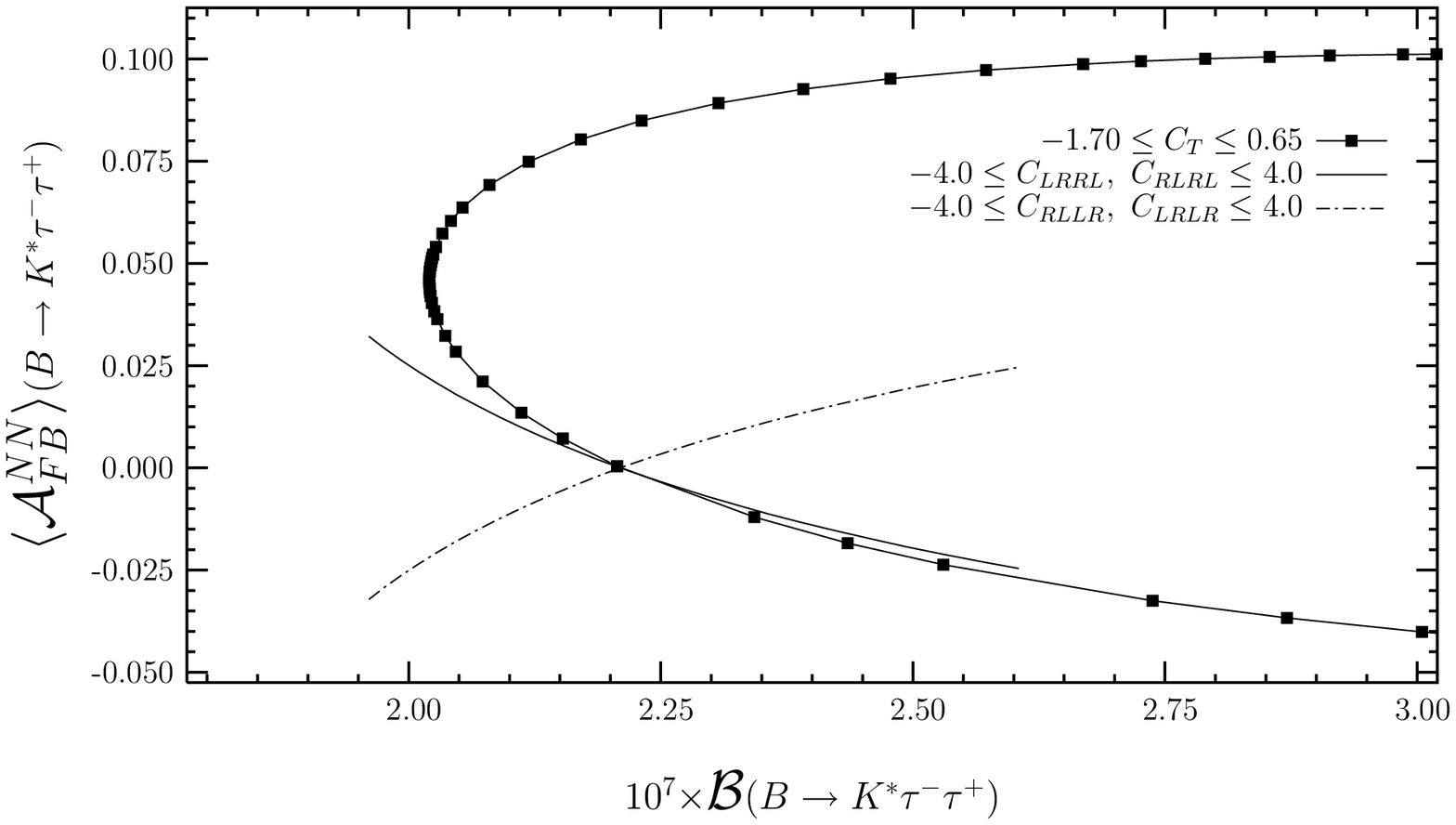}
\vskip 7.8 cm
\caption{}
\end{figure}

\begin{figure}
\vskip 2.5 cm
    \includegraphics{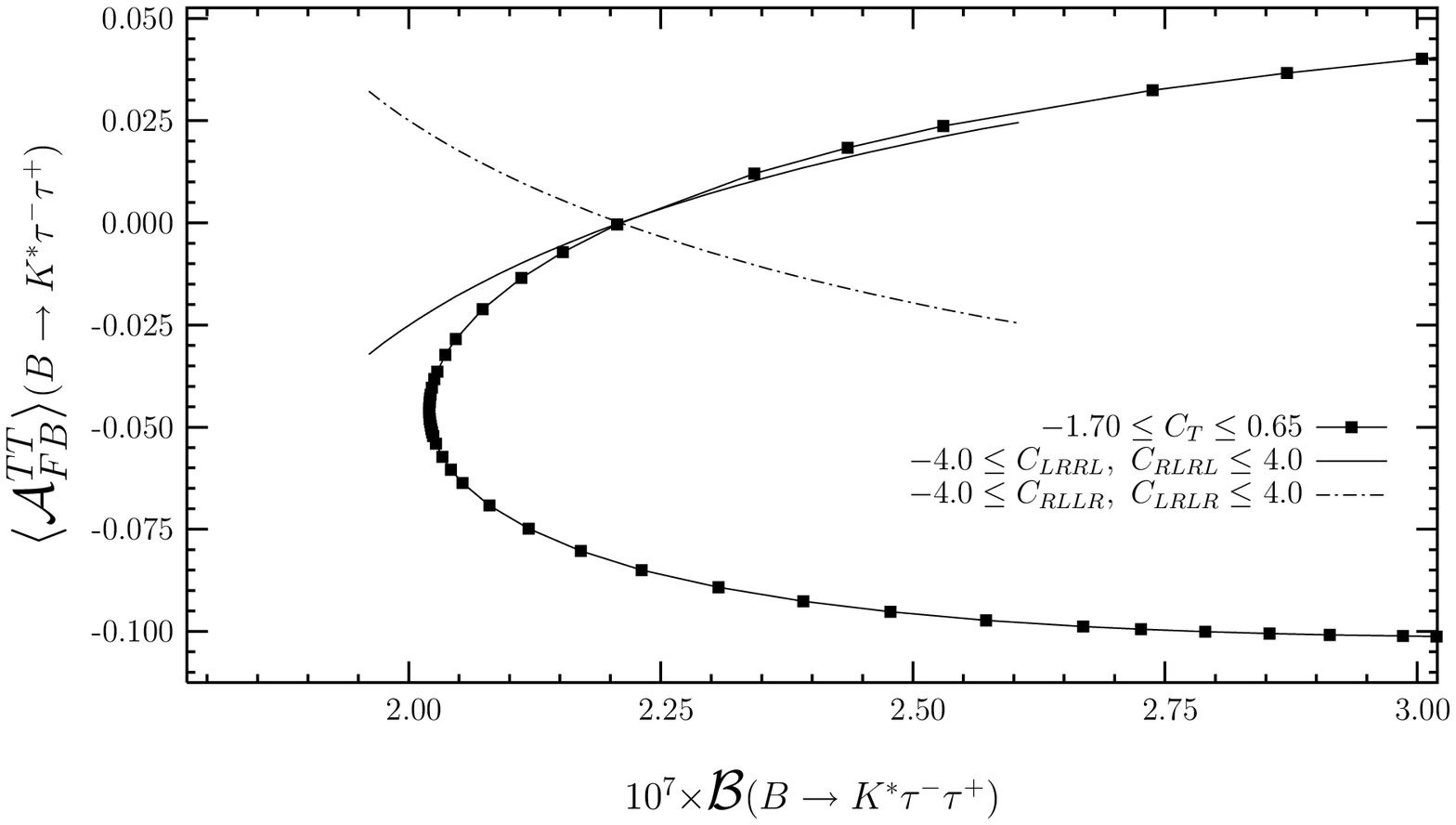}
\vskip 7.8 cm
\caption{}
\end{figure}


\newpage

\begin{thebibliography}{99}

\bibitem{R6301} 
  C. Q. Geng and C. P. Kao,
  {\it Transverse lepton polarization in $B \rar K^\ast \ell^+ \ell^-$},
  {\it Phys. Rev.} {\bf D 57} (1998) 4479 ;\\
  T. M. Aliev, M. K. \c{C}akmak and M. Savc{\i},
  {\it General analysis of lepton polarizations in rare 
  $B \rar K^\ast \ell^+ \ell^-$ decay beyond the standard model},
  {\it Nucl. Phys.} {\bf B 607} (2001) 305 [hep--ph/0009133];\\
  T. M. Aliev and M. Savc{\i},
  {\it Lepton polarization and CP--violating effects in 
  $B \rar K^\ast \tau^+ \tau^-$ decay in standard and two Higgs doublet models}, 
  {\it Phys. Lett.} {\bf B 481} (2000) 275 [hep--ph/0003188].

\bibitem{R6302} 
  T. M. Aliev, M. Savc{\i}, A. \"{O}zpineci and H. Koru,
  {\it Two Higgs doublet model and lepton polarization in the 
  $B \rar K \tau^+ \tau^-$ decay}, 
  J. Phys. G {\bf 24} (1998) 49 [hep--ph/9705222];\\
  T. M. Aliev, M. K. \c{C}akmak, A. \"{O}zpineci and M. Savc{\i},
  {\it New physics effects to the lepton polarizations in the
  $B \rar K \ell^+ \ell^-$ decay}, 
  {\it Phys. Rev.} {\bf D 64} (2001) 055007 [hep--ph/0103039].

\bibitem{R6303}
  Q.--S. Yan, C.--S. Huang, W. Liao and S.--H. Zhu,
  {\it Exclusive semileptonic rare decays $B \rar (K,K^\ast) \ell^+ \ell^-$
  in supersymmetric theories},
  {\it Phys. Rev.} {\bf D 62} (2000) 094023 [hep--ph/0004262].

\bibitem{R6304} 
  E. O. Iltan,
  {\it The exclusive $\bar{B} \rar \pi e^+ e^-$ and 
  $\bar{B} \rar \rho e^+ e^-$ decays in the two Higgs doublet model with
  flavor changing neutral currents},
  {\it Int. J. Mod. Phys.} {\bf A 14} (1999) 4365 [hep--ph/9807256];\\
  T. M. Aliev and M. Savc{\i},
  {\it Exclusive $B \rar \pi \ell^+ \ell^-$ and $B \rar \rho \ell^+ \ell^-$ 
  decays in two Higgs doublet model},
  {\it Phys. Rev.} {\bf D 60} (1999) 014005 [hep--ph/9812272];\\
  S. R. Choudhury and N. Gaur,
  {\it SUSY effects on the exclusive semi--leptonic decays 
  $B \rar \pi \ell^+ \ell^-$ and  $B \rar \rho \ell^+ \ell^-$},
  {\it Phys. Rev.} {\bf D 66} (2002) 094015 [hep--ph/0206128].

\bibitem{R6305} 
  S. R. Choudhury and N. Gaur,
  {\it Dileptonic decay of $B_s$ meson in SUSY models with large
  $\tan\beta$},
  {\it Phys. Lett.} {\bf B 451} (1999) 86 [hep--ph/9810307];\\
  P. H. Chankowski and L. Slawianowska,
  {\it $B_{d,s}^0 \rar \mu^- \mu^+$ decay in the MSSM},
  {\it Phys. Rev.} {\bf D 63} (2001) 054012 [hep--ph/0008046];\\
  A. J. Buras, P. H. Chankowski, J. Rosiek and L. Slawianowska,
  {\it $\Delta M_{d,s},B_{d,s}^0 \rar \mu^+ \mu^-$ and $B \rar X_s \gamma$
  in supersymmetry at large $\tan\beta$},
  {\it Nucl. Phys.} {\bf B 659} (2003) 3 [hep--ph/0210145];\\
  {\it Correlation between $\Delta M_s$ and $B_{s,d}^0 \rar \mu^+ \mu^-$ in
  supersymmetry at large $\tan\beta$},
  {\it Phys. Lett.} {\bf B 546}(2002) 96 [hep--ph/0207241];\\
  J. K. Mizukoshi, X. Tata and Y. Wang,
  {\it Higgs--mediated leptonic decays of $B_s$ and $B_d$ mesons as probes
  of supersymmetry},
  {\it Phys. Rev.} {\bf D 66} (2002) 115003 [hep--ph/0208078];\\
  K. S. Babu and C. F. Kolda,
  {\it Higgs--mediated $B^0 \rar \mu^+ \mu^-$ in minimal supersymmetry},
  {\it Phys. Rev. Lett.}  {\bf 84} (2000) 228;\\
  T. Ibrahim and P. Nath,
  {\it CP violation effects on $B_{s,d}^0 \rar \ell^+ \ell^-$ in
  supersymmetry at large $\tan\beta$}, 
  {\it Phys. Rev.} {\bf D 67} (2003) 016005 [hep--ph/0208142];\\
  C.--S. Huang and W. Liao,
  {\it $(g-2)_\mu$ and CP asymmetries in $B_{d,s}^0 \rar \ell^+ \ell^-$ and 
  $b \rar s \gamma$ in SUSY models},
  {\it Phys. Lett.} {\bf B 538} (2002) 301 [hep--ph/0201121];\\
  S. Baek, P. Ko and W. Y. Song,
  {\it Implications on SUSY breaking mediation mechanisms from observing 
  $B_s \rar \mu^+ \mu^-$ and the muon $(g-2)$}, 
  {\it Phys. Rev. Lett.} {\bf 89} (2002) 271801 [hep--ph/0205259];\\
  {\it SUSY breaking mediation mechanisms and $(g-2)_\mu$,
  $B \rar X_s \gamma$, $B \rar X_s \ell^+ \ell^-$ and $B_s \rar \mu^+ \mu^-$}, 
  {\it J. High Energy Phys.} {\bf 0303} (2003) 054 [hep--ph/0208112];\\
  A. Dedes and A. Pilaftsis,
  {\it Resummed effective Lagrangian for Higgs--mediated FCNC interactions
  in the CP--violating MSSM},
  {\it Phys. Rev.} {\bf D 67} (2003) 015012 [hep--ph/0209306];\\
  C.--S. Huang, W. Liao, Q.--S. Yan and S.--H. Zhu,
  {\it $B_s \rar \ell^+ \ell^-$ in a general $2HDM$ and MSSM},
  {\it Phys. Rev.} {\bf D 63} (2001) 114021 [hep--ph/0006250],
  erratum {\it ibid.} {\bf D 64} (2001) 059902;\\
  C.--S. Huang and X.--H. Wu,
  {\it $B_s \rar \mu^+ \mu^-$ and $B \rar X_s \mu^+ \mu^-$ in MSSM}, 
  {\it Nucl. Phys.} {\bf B 657} (2003) 304 [hep--ph/0212220].

\bibitem{R6306} 
  S. Rai Choudhury, N. Gaur and N. Mahajan,
  {\it Lepton polarization asymmetry in radiative dileptonic $B$ meson
  decays in MSSM},
  {\it Phys. Rev.} {\bf D 66} (2002) 054003 [hep--ph/0203041];\\
  S. R. Choudhury and N. Gaur,
  {\it Supersymmetric effects in $B_s \rar \ell^+ \ell^- \gamma$ decays},
  hep--ph/0205076;\\
  E. O. Iltan and G. Turan,
  {\it Rare radiative $B \rar \tau^+ \tau^- \gamma$ decay in the two Higgs
  doublet model},
  {\it Phys. Rev.} {\bf D 61} (2000) 034010 [hep--ph/9906502];\\
  T. M. Aliev, A. \"{O}zpineci and M. Savc{\i},
  {\it Rare $B \rar \ell^+ \ell^- \gamma$ decay and new physics effects},
  {\it Phys. Lett.} {\bf B 520} (2001) 69 [hep--ph/0105279].

\bibitem{R6307}
  A. Ali, P. Ball, L. T. Handoko and G. Hiller,
  {\it A comparative study of the decays $B \rar (K,K^\ast) \ell^+ \ell^-$
  in standard model and supersymmetric theories},
  {\it Phys. Rev.} {\bf D 61} (2000) 074024 [hep--ph/9910221].

\bibitem{R6308}
  F. Kr\"{u}ger and L. M. Sehgal,
  {\it Lepton polarization in the decays $B \rar X_s \mu^+ \mu^-$ and
  $B \rar X_s \tau^+ \tau^-$},
  {\it Phys. Lett.} {\bf B 380} (1996) 199 [hep--ph/9603237];\\
  J. L. Hewett,
  {\it Tau polarization asymmetry in $B \rar X_s \tau^+ \tau^-$},
  {\it Phys. Rev.} {\bf D 53} (1996) 4964 [hep--ph/9506289];\\
  S. Rai Choudhury, A. Gupta and N. Gaur,
  {\it Tau polarization asymmetry in $B \rar X_s \tau^+ \tau^-$ in SUSY models
  with large $\tan\beta$},
  {\it Phys. Rev.} {\bf D 60} (1999) 115004 [hep--ph/9902355].

\bibitem{R6309}
  T. M. Aliev, D. A. Demir and M. Savc{\i},
  {\it Probing the sources of CP--violation via 
  $B \rar K^\ast \ell^+ \ell^-$ decay},
  {\it Phys. Rev.} {\bf D 62} (2000) 074016 [hep--ph/9912525];\\
  T. M. Aliev, A. \"{O}zpineci, M. Savc{\i} and C. Y\"{u}ce,
  {\it T violation in $B \rar K^\ast \ell^+ \ell^-$ decay beyond
  standard model},
  {\it Phys. Rev.} D {\bf 66} (2002) 115006 [hep--ph/0208128];\\
  T. M. Aliev, A. \"{O}zpineci and M. Savc{\i},
  {\it Exclusive $B \rar K^\ast \ell^+ \ell^-$ decay with polarized $K^\ast$
  and new physics effects},
  {\it Phys. Lett.} {\bf B 511} (2001) 49 [hep--ph/0103261];\\
  {\it Fourth generation effects in rare exclusive $B \rar K^\ast \ell^+
  \ell^-$ decay},
  {\it Nucl. Phys.} {\bf B 585} (2000) 275 [hep--ph/0002061].

\bibitem{R6310} 
  T. M. Aliev, C. S. Kim and Y. G. Kim,
  {\it A systematic analysis of the exclusive $B \rar K^\ast \ell^+ \ell^-$
  decay},
  {\it Phys. Rev.} {\bf D 62} (2000) 014026 [hep--ph/9910501].
 
\bibitem{R6311}
  S. Fukae, C. S. Kim and T. Yoshikawa,
  {\it A systematic analysis of the lepton polarization asymmetries in the
  rare $B$ decay, $B \rar X_s \tau^+ \tau^-$}, 
  {\it Phys. Rev.} {\bf  D 61} (2000) 074015 [hep--ph/9908229];\\
  D. Guetta and E. Nardi,
  {\it Searching for new physics in rare $B \rar \tau$ decays},
  {\it Phys. Rev.} {\bf D 58} (1998) 012001 [hep--ph/9707371].
 
\bibitem{R6312} 
  G. Burdman,
  {\it Short distance coefficients and the vanishing of the lepton
  asymmetry in $B \rar V \ell^+ \ell^-$},
  {\it Phys. Rev.} {\bf D 57} (1998) 4254 [hep--ph/9710550].

\bibitem{R6313} 
  W. Bensalem, D. London, N. Sinha and R. Sinha,
  {\it Lepton polarization and forward--backward asymmetries in 
  $b \rar s \tau^+ \tau^-$},
  {\it Phys. Rev.} {\bf D 67} (2003) 034007 [hep--ph/0209228].

\bibitem{R6314} 
  S. R. Choudhury, N. Gaur, A. S. Cornell and G. C. Joshi,
  {\it Lepton polarization correlations in $B \rar K^\ast \tau^+ \tau^-$},
  {\it Phys. Rev.} {\bf D 68} (2003) 054016 [hep--ph/0304084].

\bibitem{R6315} T. M. Aliev, V. Bashiry and M. Savc{\i},
  {\it Double--lepton polarization asymmetries in the 
  $B \rar K \ell^+ \ell^-$ decay beyond the standard model}, 
  hep--ph/0311294.

\bibitem{R6316} 
  N. Gaur,
  {\it Lepton polarization asymmetries in 
  $B \rar X_s \tau^+ \tau^-$ in MSSM},
  hep--ph/0305242.

\bibitem{R6317}
  C. Bobeth, T. Ewerth, F. Kr\"{u}ger and J. Urban,
  {\it Analysis of neutral Higgs--boson contributions to the decays 
  $\bar{B}_s \rar \ell^+ \ell^-$ and $\bar{B} \rar K \ell^+ \ell^-$},
  {\it Phys. Rev.} {\bf D 64} (2001) 074014 [hep--ph/0104284];\\
  D. A. Demir, K. Olive and M. B. Voloshin,
  {\it The forward--backward asymmetry of $B \rar (\pi,K) \ell^+ \ell^-$:
  supersymmetry at work},
  {\it Phys. Rev.} {\bf D 66} (2002) 034015 [hep--ph/0204119]. 

\bibitem{R6318} 
  Z. Xiong and J. M. Yang,
  {\it Rare B--meson dileptonic decays in minimal supersymmetric model},
  {\it Nucl. Phys.} {\bf B 628} (2002) 193 [hep--ph/0105260];\\
  C. Bobeth, A. J. Buras, F. Kr\"{u}ger and J. Urban,
  {\it QCD corrections to $\bar{B} \rar X_{d,s} \nu \bar{\nu}$,
  $\bar{B}_{d,s} \rar \ell^+ \ell^-$, $K \rar \pi \nu \bar{\nu}$ and
  $K_L \rar \mu^+ \mu^-$ in the MSSM},
  {\it Nucl. Phys.} {\bf B 630} (2002) 87 [hep--ph/0112305];\\
  C.--S. Huang, W. Liao and Q.--S. Yan,
  {\it The promising process to distinguish supersymmetric models with large
  $\tan\beta$ from the standard model: $B \rar X_s \mu^+ \mu^-$},
  {\it Phys. Rev.} {\bf D 59} (1999) 011701 [hep--ph/9803460].

\bibitem{R6319} 
  W. Skiba and J. Kalinowski,
  {\it $B_s \rar \tau^+ \tau^-$ decay in a two--Higgs--doublet model},
  {\it Nucl. Phys.} {\bf B 404} (1993) 3;\\
  H. E. Logan and U. Nierste,
  {\it $B_{s,d} \rar \ell^+ \ell^-$ in a two--Higgs--doublet model},
  {\it Nucl. Phys.} {\bf B 586} (2000) 39 [hep--ph/0004139];\\
  Y.--B. Dai, C.--S. Huang and H.--W Huang,
  {\it $B \rar X_s \tau^+ \tau^-$ in a two--Higgs--doublet model},
  {\it Phys. Lett.} {\bf B 390} (1997) 257 [hep--ph/9607389].

\bibitem{R6320}
  Belle Collaboration, A. Ishikawa et al.,
  {\it Observation of the electroweak penguin decay
  $B \rar K^\ast \ell^+ \ell^-$},  
  {\it Phys. Rev. Lett.} {\bf 91} (2003) 261601 [hep--ex/0308044].

\bibitem{R6321} 
  BaBar Collaboration, B. Aubert et al.,
  {\it Evidence for the rare decay $B \rar K^\ast \ell^+ \ell^-$ and
  measurement of the $B \rar K  \ell^+ \ell^-$ branching fraction},
  {\it Phys. Rev. Lett.} {\bf 91} (2003) 221802 [hep--ex/0308042].

\bibitem{R6322} A. Ali, E. Lunghi, C. Greub and G. Hiller,
  {\it Improved model--independent analysis of semileptonic and radiative
  rare $B$ decays},
  {\it Phys. Rev.} {\bf D 66} (2002) 034002 [hep--ph/0112300];\\
  H. M. Asatrian, K. Bieri, C. Greub and A. Hovhannisyan,
  {\it NNLL corrections to the angular distribution and to the
  forward--backward asymmetries in $B \rar X_s \ell^+ \ell^-$},
  {\it Phys. Rev.} {\bf D 66} (2002) 094013 [hep--ph/0209006];\\
  A. Ghinculov, T. Hurth, G. Isidori and Y.P. Yao,
  {\it Forward--backward asymmetry in $B \rar X_s \ell^+ \ell^-$ at the
  NNLL level},
  {\it Nucl. Phys.} {\bf B 648} (2003) 254 [hep--ph/0208088].  

\bibitem{R6323}
  S. R. Choudhury, N. Gaur, A. S. Cornell and G. C. Joshi,
  {\it Supersymmetric effects on forward--backward asymmetries of
  $B \rar K \ell^+ \ell^-$},
  {\it Phys. Rev.} {\bf D 69} (2004) 054018 [hep--ph/0307276];\\
  A. S. Cornell and N. Gaur, 
  {\it The forward--backward asymmetries of $B \rar X_s \tau^+ \tau^-$ in the
  MSSM},
  {\it J. High Energy Phys.} {\bf 0309} (2003) 030 [hep--ph/0308132].

\bibitem{R6324}
  P. Ball and V. M. Braun,
  {\it Exclusive semileptonic and rare $B$ meson decays in QCD},
  {\it Phys. Rev.} {\bf D 58} (1998) 094016 [hep--ph/9805422].

\bibitem{R6325}  T. M. Aliev, A. \"{O}zpineci and M. Savc{\i},
  {\it Rare $B \rar K^\ast \ell^+ \ell^-$ decay in light cone QCD}, 
  {\it Phys. Rev.} {\bf D 56} (1997) 4260 [hep--ph/9612480].

\bibitem{R6326} 
  V. Halyo,
  {\it New BaBar results on rare leptonic $B$ decays},
  hep--ex/0207010.

\end{thebibliography}
\end{document}